\documentclass[aip,jcp,reprint,noshowkeys]{revtex4-1}
\usepackage{graphicx,dcolumn,bm,xcolor,microtype,multirow,amscd,amsmath,amssymb,amsfonts,physics,mhchem,longtable,wrapfig,txfonts}

\usepackage{natbib}
\usepackage[extra]{tipa}
\AtBeginDocument{\nocite{achemso-control}}

\usepackage{hyperref}
\hypersetup{
    colorlinks=true,
    linkcolor=blue,
    filecolor=blue,
    urlcolor=blue,
    citecolor=blue
}
\newcommand{\ie}{\textit{i.e}}
\newcommand{\eg}{\textit{e.g}}

\definecolor{darkgreen}{HTML}{009900}
\usepackage[normalem]{ulem}
\newcommand{\titou}[1]{\textcolor{black}{#1}}

\newcommand{\mc}{\multicolumn}
\newcommand{\fnm}{\footnotemark}
\newcommand{\fnt}{\footnotetext}
\newcommand{\tabc}[1]{\multicolumn{1}{c}{#1}}
\newcommand{\SI}{\textcolor{blue}{supporting information}}

\newcommand{\HF}{\text{HF}}
\newcommand{\KS}{\text{KS}}
\newcommand{\GOWO}{$G_0W_0$}	
\newcommand{\evGW}{ev$GW$}	
\newcommand{\qsGW}{qs$GW$}	
\newcommand{\GW}{$GW$}		
\newcommand{\Hxc}{\text{Hxc}}
\newcommand{\Hx}{\text{Hx}}
\newcommand{\xc}{\text{xc}}

\newcommand{\Bas}{\mathcal{B}}
\newcommand{\cD}{\mathcal{D}}
\newcommand{\Ne}{N}
\newcommand{\Nbas}{N_\text{bas}}
\newcommand{\Nocc}{N_\text{occ}}
\newcommand{\Nvirt}{N_\text{virt}}
\newcommand{\Ngrid}{N_\text{grid}}

\newcommand{\IP}{\text{IP}}
\newcommand{\EA}{\text{EA}}
\newcommand{\CBS}{\text{CBS}}

\newcommand{\PBEO}{\text{PBE0}}
\newcommand{\srLDA}{\text{srLDA}}
\newcommand{\srPBE}{\text{srPBE}}
\newcommand{\be}[2]{\Bar{\varepsilon}_{#1}^{#2}}

\newcommand{\e}[1]{\epsilon_{#1}}

\newcommand{\eGOWO}[1]{\epsilon^\text{\GOWO}_{#1}}
\newcommand{\beGOWO}[1]{\Bar{\epsilon}^\text{\GOWO}_{#1}}
\newcommand{\Om}[1]{\Omega_{#1}}

\newcommand{\HOMO}{\text{HOMO}}
\newcommand{\LUMO}{\text{LUMO}}

\newcommand{\G}[2]{G_{#1}^{#2}}
\newcommand{\Gs}[1]{G_\text{f}^{#1}}
\newcommand{\F}[2]{F_{#1}^{#2}}
\newcommand{\Po}[2]{P_{#1}^{#2}}
\newcommand{\W}[2]{W_{#1}^{#2}}

\newcommand{\vc}[2]{\varv_{#1}^{#2}}
\newcommand{\pot}[2]{v_{#1}^{#2}}
\newcommand{\Pot}[2]{V_{#1}^{#2}}
\newcommand{\bpot}[2]{\Bar{v}_{#1}^{#2}}
\newcommand{\bPot}[2]{\Bar{V}_{#1}^{#2}}
\newcommand{\Sig}[2]{\Sigma_{#1}^{#2}}
\newcommand{\bSig}[2]{\Bar{\Sigma}_{#1}^{#2}}
\newcommand{\Z}[1]{Z_{#1}}
\newcommand{\Gam}[2]{\Gamma_{#1}^{#2}}

\newcommand{\vne}{v_\text{ne}}
\newcommand{\hT}{\Hat{T}}
\newcommand{\hWee}[1]{\Hat{W}_\text{ee}^{#1}}
\newcommand{\MO}[1]{\phi_{#1}}
\newcommand{\stat}[1]{\underset{#1}{\text{stat}}}

\newcommand{\bA}{\boldsymbol{A}}
\newcommand{\bB}{\boldsymbol{B}}
\newcommand{\bX}{\boldsymbol{X}}
\newcommand{\bY}{\boldsymbol{Y}}

\newcommand{\br}[1]{\mathbf{r}_{#1}}
\newcommand{\dbr}[1]{d\br{#1}}

\newcommand{\n}[2]{n_{#1}^{#2}}
\newcommand{\E}[2]{E_{#1}^{#2}}
\newcommand{\bE}[2]{\Bar{E}_{#1}^{#2}}

\newcommand{\wf}[2]{\Psi_{#1}^{#2}}

\newcommand{\rsmu}[2]{\mu_{#1}^{#2}}

\newcommand{\kcal}{kcal/mol}

\newcommand{\ISCD}{Institut des Sciences du Calcul et des Donn\'ees, Sorbonne Universit\'e, Paris, France}
\newcommand{\LCPQ}{Laboratoire de Chimie et Physique Quantiques (UMR 5626), Universit\'e de Toulouse, CNRS, UPS, France}
\newcommand{\LCT}{Laboratoire de Chimie Th\'eorique (UMR 7616), Sorbonne Universit\'e, CNRS, Paris, France}
\newcommand{\IUF}{Institut Universitaire de France, Paris, France}

\begin{document}	

\title{A Density-Based Basis-Set Incompleteness Correction for GW Methods}

\author{Pierre-Fran\c{c}ois Loos}
\email[Corresponding author: ]{loos@irsamc.ups-tlse.fr}
\affiliation{\LCPQ}
\author{Barth\'el\'emy Pradines}
\affiliation{\LCT}
\affiliation{\ISCD}
\author{Anthony Scemama}
\affiliation{\LCPQ}
\author{Emmanuel Giner}
\affiliation{\LCT}
\author{Julien Toulouse}
\email[Corresponding author: ]{toulouse@lct.jussieu.fr}
\affiliation{\LCT}
\affiliation{\IUF}

\begin{abstract}
\begin{wrapfigure}[12]{o}[-1.25cm]{0.4\linewidth}
	\centering
  \includegraphics[width=\linewidth]{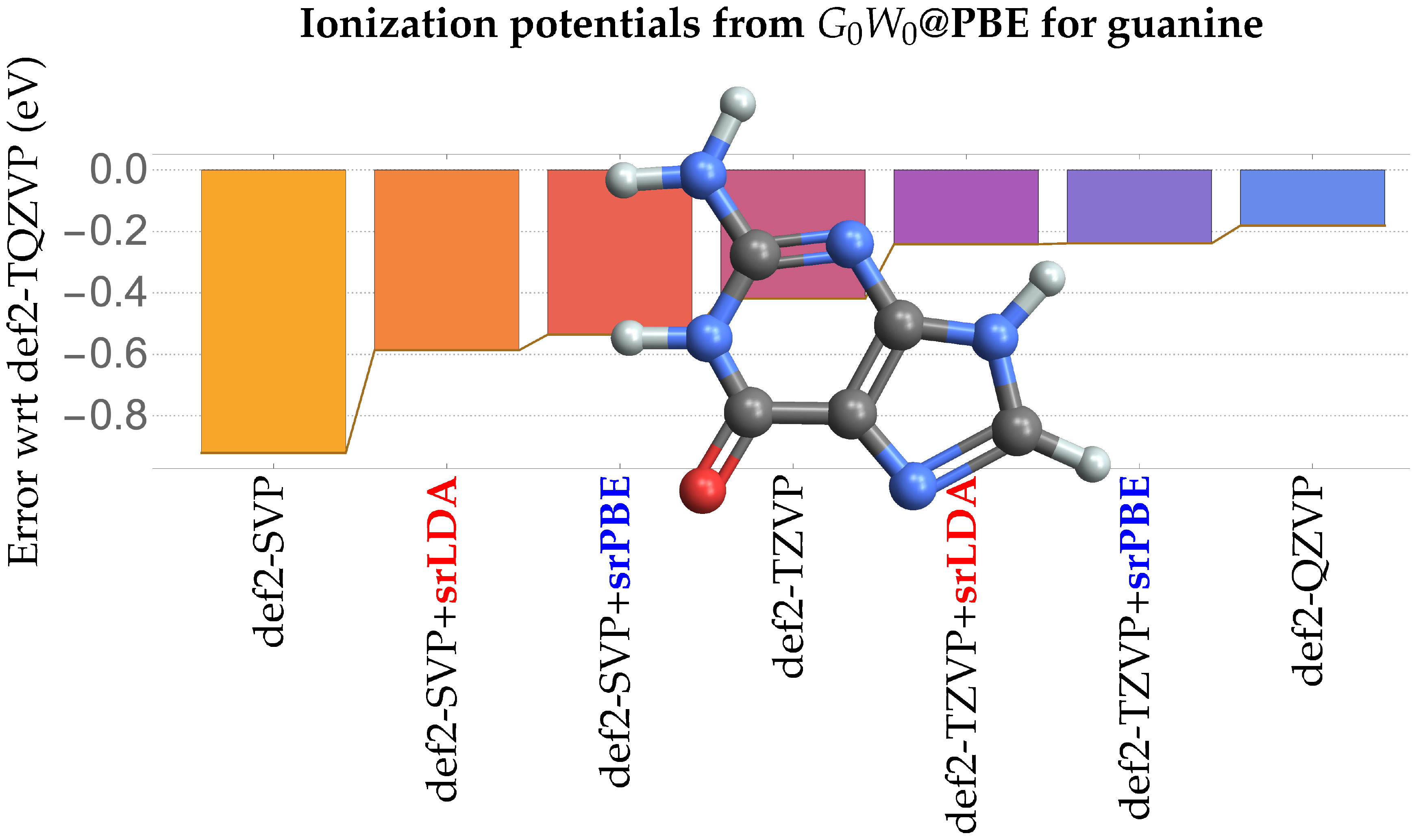}
\end{wrapfigure}
Similar to other electron correlation methods, many-body perturbation theory methods based on Green functions, such as the so-called $GW$ approximation, suffer from the usual slow convergence of energetic properties with respect to the size of the one-electron basis set.
This displeasing feature is due to the lack of explicit electron-electron terms modeling the infamous Kato electron-electron cusp and the correlation Coulomb hole around it.
Here, we propose a computationally efficient density-based basis-set correction based on short-range correlation density functionals which significantly speeds up the convergence of energetics towards the complete basis set limit.
The performance of this density-based correction is illustrated by computing the ionization potentials of the twenty smallest atoms and molecules of the GW100 test set at the perturbative $GW$ (or $G_0W_0$) level using increasingly large basis sets.
We also compute the ionization potentials of the five canonical nucleobases (adenine, cytosine, thymine, guanine, and uracil) and show that, here again, a significant improvement is obtained.
\end{abstract}

\maketitle

%%%%%%%%%%%%%%%%%%%%%%%%
\section{Introduction}
\label{sec:intro}
%%%%%%%%%%%%%%%%%%%%%%%%
The purpose of many-body perturbation theory (MBPT) based on Green functions is to solve the formidable many-body problem by adding the electron-electron Coulomb interaction perturbatively starting from an independent-particle model. \cite{Martin_2016} 
In this approach, the \textit{screening} of the Coulomb interaction is an essential quantity. \cite{Aryasetiawan_1998, Onida_2002, Reining_2017} 

The so-called {\GW} approximation is the workhorse of MBPT and has a long and successful history in the calculation of the electronic structure of solids. \cite{Aryasetiawan_1998, Onida_2002, Reining_2017} 
{\GW} is getting increasingly popular in molecular systems \cite{Blase_2011, Faber_2011, Bruneval_2012, Bruneval_2015, Bruneval_2016, Bruneval_2016a, Boulanger_2014, Blase_2016, Li_2017, Hung_2016, Hung_2017, vanSetten_2015, vanSetten_2018, Ou_2016, Ou_2018, Faber_2014} thanks to efficient implementation relying on plane waves \cite{Marini_2009, Deslippe_2012, Maggio_2017} or local basis functions. \cite{Blase_2011, Blase_2018, Bruneval_2016, vanSetten_2013, Kaplan_2015, Kaplan_2016, Krause_2017, Caruso_2012, Caruso_2013, Caruso_2013a, Caruso_2013b}
The {\GW} approximation stems from the acclaimed Hedin's equations \cite{Hedin_1965} 
\begin{subequations}
\begin{align}
	\label{eq:G}
	& \G{}{}(12) = \G{0}{}(12) + \int \G{0}{}(13) \Sig{}{}(34) \G{}{}(42) d(34),
	\\
	\label{eq:Gamma}
	& \Gam{}{}(123) = \delta(12) \delta(13) 
	\notag 
	\\
	& \qquad \qquad		+ \int \fdv{\Sig{}{}(12)}{\G{}{}(45)} \G{}{}(46) G(75) \Gam{}{}(673) d(4567),
	\\
	\label{eq:P}
	& \Po{}{}(12) = - i \int G(13)  G(41) \Gam{}{}(342) d(34),
	\\
	\label{eq:W}
	& \W{}{}(12) = \vc{}{}(12) + \int \vc{}{}(13) \Po{}{}(34) \W{}{}(42) d(34),
	\\
	\label{eq:Sig}
	& \Sig{}{}(12) = i \int \G{}{}(13) \W{}{}(14) \Gam{}{}(324) d(34),
\end{align}
\end{subequations}
which connects the Green function $\G{}{}$, its non-interacting version $\G{0}{}$, the irreducible vertex function $\Gam{}{}$, the irreducible polarizability $\Po{}{}$, the dynamically-screened Coulomb interaction $\W{}{}$ 
and the self-energy $\Sig{}{}$, where $\vc{}{}$ is the bare Coulomb interaction, $\delta(12)$ is the Dirac delta function \cite{NISTbook} and $1$ is a composite coordinate gathering space, spin, and time variables $(\br{1},\sigma_1,t_1)$.
Within the {\GW} approximation, one bypasses the calculation of the vertex corrections by setting 
\begin{equation}
	\label{eq:GW}
	\Gam{}{}(123) \stackrel{GW}{\approx} \delta(12) \delta(13).
\end{equation}
Depending on the degree of self-consistency one is willing to perform, there exists several types of {\GW} calculations. \cite{Loos_2018}
The simplest and most popular variant of {\GW} is perturbative {\GW} (or {\GOWO}). \cite{Hybertsen_1985a, Hybertsen_1986}
Although obviously starting-point dependent, \cite{Bruneval_2013, Jacquemin_2016, Gui_2018} it has been widely used in the literature to study solids, atoms, and molecules. \cite{Bruneval_2012, Bruneval_2013, vanSetten_2015, vanSetten_2018}
For finite systems such as atoms and molecules, partially \cite{Hybertsen_1986, Shishkin_2007, Blase_2011, Faber_2011} or fully self-consistent \cite{Caruso_2012, Caruso_2013, Caruso_2013a, Caruso_2013b} {\GW} methods have shown great promise. \cite{Ke_2011, Blase_2011, Faber_2011, Koval_2014, Hung_2016, Blase_2018, Jacquemin_2017} 

Similar to other electron correlation methods, MBPT methods suffer from the usual slow convergence of energetic properties with respect to the size of the one-electron basis set.
This can be tracked down to the lack of explicit electron-electron terms modeling the infamous electron-electron coalescence point (also known as Kato cusp \cite{Kato_1957}) and, more specifically, the Coulomb correlation hole around it.
Pioneered by Hylleraas \cite{Hylleraas_1929} in the 1930's and popularized in the 1990's by Kutzelnigg and coworkers \cite{Kutzelnigg_1985, Noga_1994, Kutzelnigg_1991} (and subsequently others \cite{Kong_2012, Hattig_2012, Tenno_2012a, Tenno_2012b, Gruneis_2017}), the so-called F12 methods overcome this slow convergence by employing geminal basis functions that closely resemble the correlation holes in electronic wave functions.
F12 methods are now routinely employed in computational chemistry and provide robust tools for electronic structure calculations where small basis sets may be used to obtain near complete basis set (CBS) limit accuracy. \cite{Tew_2007}

The basis-set correction presented here follow a different route, and relies on the range-separated density-functional theory (RS-DFT) formalism to capture, thanks to a short-range correlation functional, the missing part of the short-range correlation effects. \cite{Giner_2018}
As shown in recent studies on both ground- and excited-state properties, \cite{Loos_2019, Giner_2019} similar to F12 methods, it significantly speeds up the convergence of energetics towards the CBS limit while avoiding the usage of the large auxiliary basis sets that are used in F12 methods to avoid the numerous three- and four-electron integrals. \cite{Kong_2012, Hattig_2012, Tenno_2012a, Tenno_2012b, Gruneis_2017, Barca_2016, Barca_2017, Barca_2018}

Explicitly correlated F12 correction schemes have been derived for second-order Green function methods (GF2) \cite{SzaboBook, Casida_1989, Casida_1991, Stefanucci_2013, Ortiz_2013, Phillips_2014, Phillips_2015, Rusakov_2014, Rusakov_2016, Hirata_2015, Hirata_2017, Loos_2018} by Ten-no and coworkers \cite{Ohnishi_2016, Johnson_2018} and Valeev and coworkers. \cite{Pavosevic_2017, Teke_2019}
However, to the best of our knowledge, a F12-based correction for {\GW} has not been designed yet.

In the present manuscript, we illustrate the performance of the density-based basis-set correction developed in Refs.~\onlinecite{Giner_2018, Loos_2019, Giner_2019} on ionization potentials obtained within {\GOWO}. 
Note that the present basis-set correction can be straightforwardly applied to other properties (\eg, electron affinities and fundamental gaps), as well as other flavors of (self-consistent) {\GW} or Green function-based methods, such as GF2 (and its higher-order variants). 

The paper is organized as follows. 
In Sec.~\ref{sec:theory}, we provide details about the theory behind the present basis-set correction and its adaptation to {\GW} methods.
Results for a large collection of molecular systems are reported and discussed in Sec.~\ref{sec:results}. 
Finally, we draw our conclusions in Sec.~\ref{sec:conclusion}. 
Unless otherwise stated, atomic units are used throughout.

%%%%%%%%%%%%%%%%%%%%%%%%%%%%%%%%
\section{Theory}
\label{sec:theory}
%%%%%%%%%%%%%%%%%%%%%%%%%%%%%%%%

%%%%%%%%%%%%%%%%%%%%%%%%%%%%%%%%
\subsection{MBPT with DFT basis-set correction}
%%%%%%%%%%%%%%%%%%%%%%%%%%%%%%%%

Following Ref.~\onlinecite{Giner_2018}, we start by defining, for a $\Ne$-electron system with nuclei-electron potential $\vne(\br{})$, the approximate ground-state energy for one-electron densities $\n{}{}$ which are ``representable'' in a finite basis set $\Bas$
\begin{equation}
	\E{0}{\Bas} = \min_{\n{}{} \in \cD^\Bas} \qty{ \F{}{}[n] + \int \vne(\br{}) \n{}{}(\br{}) \dbr{} },
\label{eq:E0B}
\end{equation}
where $\cD^\Bas$ is the set of $\Ne$-representable densities which can be extracted from a wave function $\wf{}{\Bas}$ expandable in the Hilbert space generated by $\Bas$. 
In this expression, 
\begin{equation}
	\F{}{}[n] = \min_{\wf{}{} \rightsquigarrow \n{}{}} \mel*{\wf{}{}}{\hT + \hWee{}}{\wf{}{}}
\end{equation}
is the exact Levy-Lieb universal density functional, \cite{Levy_1979, Levy_1982, Lieb_1983} where the notation $\wf{}{} \rightsquigarrow \n{}{}$ in Eq.~\eqref{eq:E0B} states that $\wf{}{}$ yields the one-electron density $\n{}{}$.
$\hT$ and $\hWee{}$ are the kinetic and electron-electron interaction operators.
The exact Levy-Lieb universal density functional is then decomposed as
\begin{equation}
	\F{}{}[\n{}{}] = \F{}{\Bas}[\n{}{}] + \bE{}{\Bas}[\n{}{}],
\label{eq:Fn}
\end{equation}
where $\F{}{\Bas}[\n{}{}]$ is the Levy-Lieb density functional with wave functions $\wf{}{\Bas}$ expandable in the Hilbert space generated by $\Bas$
\begin{equation}
	\F{}{\Bas}[\n{}{}] = \min_{\wf{}{\Bas} \rightsquigarrow \n{}{}} \mel*{\wf{}{\Bas}}{ \hT + \hWee{}}{\wf{}{\Bas}},
\end{equation}
and $\bE{}{\Bas}[\n{}{}]$ is the complementary basis-correction density functional. \cite{Giner_2018}
In the present work, instead of using wave-function methods for calculating $\F{}{\Bas}[\n{}{}]$, we use Green-function methods. 
We assume that there exists a functional $\Omega^\Bas[\G{}{\Bas}]$ of $\Ne$-representable one-electron Green functions $\G{}{\Bas}(\br{},\br{}',\omega)$ representable in the basis set $\Bas$ and yielding the density $\n{}{}$ which gives $\F{}{\Bas}[\n{}{}]$ at a stationary point
\begin{equation}
	\F{}{\Bas}[\n{}{}] = \stat{\G{}{\Bas} \rightsquigarrow \n{}{}} \Omega^\Bas[\G{}{\Bas}].
\label{eq:FBn}
\end{equation}
The reason why we use a stationary condition rather than a minimization condition is that only a stationary property is generally known for functionals of the Green function. 
For example, we can choose for $\Omega^\Bas[\G{}{}]$ a Klein-like energy functional (see, \eg, Refs.~\onlinecite{Stefanucci_2013, Martin_2016, Dahlen_2005, Dahlen_2005a, Dahlen_2006})
\begin{equation}
	\Omega^\Bas[\G{}{}] = \Tr[\ln( - \G{}{} ) ] - \Tr[ (\Gs{\Bas})^{-1} \G{}{} - 1 ]  + \Phi_\Hxc^\Bas[\G{}{}],
\label{eq:OmegaB}
\end{equation}
where $(\Gs{\Bas})^{-1}$ is the projection into $\Bas$ of the inverse free-particle Green function 
\begin{equation}
	(\Gs{})^{-1}(\br{},\br{}',\omega)= \qty(\omega + \frac{\nabla^2_{\br{}}}{2} ) \delta(\br{}-\br{}'),
\end{equation}
and we have introduced the trace 
\begin{equation}
	\Tr[A B] = \int_{-\infty}^{+\infty} \frac{d\omega}{2\pi i} e^{i \omega 0^+} \iint A(\br{},\br{}',\omega) B(\br{}',\br{}{},\omega)  \dbr{} \dbr{}'.
\end{equation} 
In Eq.~\eqref{eq:OmegaB}, $\Phi_\Hxc^\Bas[\G{}{}]$ is a Hartree-exchange-correlation ($\Hxc$) functional of the Green function such that its functional derivatives yields the $\Hxc$ self-energy in the basis
\begin{equation}
	\fdv{\Phi_\Hxc^\Bas[\G{}{}]}{\G{}{}(\br{},\br{}',\omega)} = \Sig{\Hxc}{\Bas}[\G{}{}](\br{},\br{}',\omega).
\end{equation}
Inserting Eqs.~\eqref{eq:Fn} and \eqref{eq:FBn} into Eq.~\eqref{eq:E0B}, we finally arrive at
\begin{equation}
	\E{0}{\Bas} = \stat{\G{}{\Bas}} \qty{ \Omega^\Bas[\G{}{\Bas}] + \int \vne(\br{}) \n{\G{}{\Bas}}{}(\br{}) \dbr{} + \bE{}{\Bas}[\n{\G{}{\Bas}}{}] },
\label{eq:E0BGB}
\end{equation}
where the stationary point is searched over $\Ne$-representable one-electron Green functions $\G{}{\Bas}(\br{},\br{}',\omega)$ representable in the basis set $\Bas$.

The stationary condition from Eq.~\eqref{eq:E0BGB} is
\begin{multline}
	\fdv{}{\G{}{\Bas}} \Bigg( \Omega^\Bas[\G{}{\Bas}] + \int \vne(\br{}) \n{\G{}{\Bas}}{}(\br{}) \dbr{} + \bE{}{\Bas}[\n{\G{}{\Bas}}{}] 
	\\
	- \lambda \int \n{\G{}{\Bas}}{}(\br{}) \dbr{} \Bigg) = 0,
\label{eq:stat}
\end{multline}
where $\lambda$ is the chemical potential (enforcing the electron number). 
It leads the following Dyson equation
\begin{equation}
	(\G{}{\Bas})^{-1} = (\G{0}{\Bas})^{-1}- \Sig{\Hxc}{\Bas}[\G{}{\Bas}]- \bSig{}{\Bas}[\n{\G{}{\Bas}}{}],
\label{eq:Dyson}
\end{equation}
where $(\G{0}{\Bas})^{-1}$ is the basis projection of the inverse non-interacting Green function with potential $\vne(\br{})$, \ie,
\begin{equation}
	(\G{0}{})^{-1}(\br{},\br{}',\omega)= \qty(\omega + \frac{\nabla_{\br{}}^2}{2} - \vne(\br{}) + \lambda) \delta(\br{}-\br{}'),
\end{equation}
and $\bSig{}{\Bas}$ is a frequency-independent local self-energy coming from the functional derivative of the complementary basis-correction density functional
\begin{equation}
	\bSig{}{\Bas}[\n{}{}](\br{},\br{}') = \bpot{}{\Bas}[\n{}{}](\br{}) \delta(\br{}-\br{}'),
\end{equation}
with $\bpot{}{\Bas}[\n{}{}](\br{}) = \delta \bE{}{\Bas}[\n{}{}] / \delta \n{}{}(\br{})$. 
This is found from Eq.~\eqref{eq:stat} by using the chain rule,
\begin{equation}
	\fdv{\bE{}{\Bas}[\n{}{}]}{\G{}{}(\br{},\br{}',\omega)} = \int \fdv{\bE{}{\Bas}[\n{}{}]}{\n{}{}(\br{}'')}  \fdv{\n{}{}(\br{}'')}{\G{}{}(\br{},\br{}',\omega)} \dbr{}'',
\end{equation}
and 
\begin{equation}
	\n{}{}(\br{}) = \int_{-\infty}^{+\infty} \frac{d\omega}{2\pi i} e^{i \omega 0^+} \G{}{}(\br{},\br{},\omega).
\end{equation}
The solution of the Dyson equation \eqref{eq:Dyson} gives the Green function $\G{}{\Bas}(\br{},\br{}',\omega)$ which is not exact (even using the exact complementary basis-correction density functional $\bSig{}{\Bas}[\n{}{}]$) but should converge more rapidly with the basis set thanks to the presence of the basis-set correction $\bSig{}{\Bas}$. 
Of course, in the CBS limit, the basis-set correction vanishes and the Green function becomes exact, \ie,
\begin{align}
	\label{eq:limSig}
	\lim_{\Bas \to \CBS} \bSig{}{\Bas} & = 0,
	 &
	\lim_{\Bas \to \CBS} \G{}{\Bas} & = \G{}{}.
\end{align}

The Dyson equation \eqref{eq:Dyson} can also be written with an arbitrary reference
\begin{equation}
	(\G{}{\Bas})^{-1} = (\G{\text{ref}}{\Bas})^{-1} - \qty( \Sig{\Hxc}{\Bas}[\G{}{\Bas}]- \Sig{\text{ref}}{\Bas} )  - \bSig{}{\Bas}[\n{\G{}{\Bas}}{}],
\end{equation}
where $(\G{\text{ref}}{\Bas})^{-1} = (\G{0}{\Bas})^{-1} - \Sig{\text{ref}}{\Bas}$. 
For example, if the reference is Hartree-Fock ($\HF$), $\Sig{\text{ref}}{\Bas}(\br{},\br{}') = \Sig{\Hx}{\Bas}(\br{},\br{}')$ is the $\HF$ nonlocal self-energy, and if the reference is Kohn-Sham ($\KS$), $\Sig{\text{ref}}{\Bas}(\br{},\br{}') = \pot{\Hxc}{\Bas}(\br{}) \delta(\br{}-\br{}')$ is the local $\Hxc$ potential.

\titou{Note that the present basis-set correction is applicable to any approximation of the self-energy (irrespectively of the diagrams included) without altering the CBS limit of such methods.
Consequently, it can be applied, for example, to GF2 methods (also known as second Born approximation \cite{Stefanucci_2013} in the condensed-matter community) or higher orders. \cite{SzaboBook, Casida_1989, Casida_1991, Stefanucci_2013, Ortiz_2013, Phillips_2014, Phillips_2015, Rusakov_2014, Rusakov_2016, Hirata_2015, Hirata_2017, Loos_2018}
Note, however, that the basis-set correction is optimal for the \textit{exact} self-energy within a given basis set, since it corrects only for the basis-set error and not for the chosen approximate form of the self-energy within the basis set.}

%%%%%%%%%%%%%%%%%%%%%%%%%%%%%%%%
\subsection{The {\GW} Approximation}
%%%%%%%%%%%%%%%%%%%%%%%%%%%%%%%%

In this subsection, we provide the minimal set of equations required to describe {\GOWO}.
More details can be found, for example, in Refs.~\citenum{vanSetten_2013, Kaplan_2016, Bruneval_2016}.
For the sake of simplicity, we only give the equations for closed-shell systems with a $\KS$ single-particle reference (with a local potential).
The one-electron energies $\e{p}$ and their corresponding (real-valued) orbitals $\MO{p}(\br{})$ (which defines the basis set $\Bas$) are then the $\KS$ orbitals and their orbital energies.

Within the {\GW} approximation, the correlation part of the self-energy reads 
\begin{equation}
\label{eq:SigC}
\begin{split}
	\Sig{\text{c},p}{\Bas}(\omega)
	& = \mel*{\MO{p}}{\Sig{\text{c}}{\Bas}(\omega)}{\MO{p}}
	\\
	& = 2 \sum_{i}^{\Nocc} \sum_{m} \frac{[pi|m]^2}{\omega - \e{i} + \Om{m} - i \eta}
	\\
	& + 2 \sum_{a}^{\Nvirt} \sum_{m} \frac{[pa|m]^2}{\omega - \e{a} - \Om{m} + i \eta},
\end{split}
\end{equation}
\titou{where $i$ runs over the $\Nocc$ occupied orbitals, $a$ runs over the $\Nvirt$ virtual orbitals, $m$ labels excited states (see below)}, and $\eta$ is a positive infinitesimal.
The screened two-electron integrals
\begin{equation}
	[pq|m] =  \sum_{i}^{\Nocc} \sum_{a}^{\Nvirt}  (pq|ia) (\bX_m+\bY_m)_{ia}
\end{equation}
are obtained via the contraction of the bare two-electron integrals \cite{Gill_1994} 
\begin{equation}
	(pq|rs) = \iint \frac{\MO{p}(\br{}) \MO{q}(\br{}) \MO{r}(\br{}') \MO{s}(\br{}')}{\abs*{\br{} - \br{}'}} \dbr{} \dbr{}',
\end{equation}
and the transition densities $(\bX_m+\bY_m)_{ia}$ originating from a (direct) random-phase approximation (RPA) calculation \cite{Casida_1995, Dreuw_2005}
\begin{equation}
\label{eq:LR}
	\begin{pmatrix}
		\bA	&	\bB	\\
		-\bB	&	-\bA	\\
	\end{pmatrix}
	\begin{pmatrix}
		\bX_m	\\
		\bY_m	\\
	\end{pmatrix}
	=
	\Om{m}
	\begin{pmatrix}
		\bX_m	\\
		\bY_m	\\
	\end{pmatrix},
\end{equation}
with
\begin{align}
\label{eq:RPA}
	A_{ia,jb} & = \delta_{ij} \delta_{ab} (\e{a} - \e{i}) + 2 (ia|bj),
	&
	B_{ia,jb} & = 2 (ia|jb),
\end{align}
and $\delta_{pq}$ is the Kronecker delta. \cite{NISTbook}
Equation \eqref{eq:LR} also provides the RPA neutral excitation energies $\Om{m}$ which correspond to the poles of the screened Coulomb interaction $\W{}{}(\omega)$.

The {\GOWO} quasiparticle energies $\eGOWO{p}$ are provided by the solution of the (non-linear) quasiparticle equation \cite{Hybertsen_1985a, vanSetten_2013, Veril_2018}
\begin{equation}
\label{eq:QP-G0W0}
	\omega =  \e{p} - \Pot{\xc,p}{\Bas} + \Sig{\text{x},p}{\Bas} + \Re[\Sig{\text{c},p}{\Bas}(\omega)].
\end{equation}
with the largest renormalization weight (or factor)
\begin{equation}
\label{eq:Z}
	\Z{p} = \qty[ 1 - \left. \pdv{\Re[\Sig{\text{c},p}{\Bas}(\omega)]}{\omega} \right|_{\omega = \e{p}}]^{-1}.
\end{equation}
Because of sum rules, \cite{Martin_1959, Baym_1961, Baym_1962, vonBarth_1996} the other solutions, known as satellites, share the remaining weight.
In Eq.~\eqref{eq:QP-G0W0}, $\Sig{\text{x},p}{\Bas} = \mel*{\MO{p}}{\Sig{\text{x}}{\Bas}}{\MO{p}}$ is the (static) HF exchange part of the self-energy and
\begin{equation}
	\Pot{\xc,p}{\Bas} = \int \MO{p}(\br{}) \pot{\xc}{\Bas}(\br{}) \MO{p}(\br{}) \dbr{},
\end{equation}
where $\pot{\xc}{\Bas}(\br{})$ is the KS exchange-correlation potential.
In particular, the ionization potential (IP) and electron affinity (EA) are extracted thanks to the following relationships: \cite{SzaboBook}
\begin{align}
  \IP   & = -\eGOWO{\HOMO},
	&
  \EA   & = -\eGOWO{\LUMO},
\end{align}
where $\eGOWO{\HOMO}$ and $\eGOWO{\LUMO}$ are the HOMO and LUMO quasiparticle energies, respectively.

%%%%%%%%%%%%%%%%%%%%%%%%
\subsection{Basis-set correction}
\label{sec:BSC}
%%%%%%%%%%%%%%%%%%%%%%%%
\titou{The fundamental idea behind the present basis-set correction is to recognize that the singular two-electron Coulomb interaction $\abs*{\br{} - \br{}'}^{-1}$ projected in a finite basis $\Bas$ is a finite, non-divergent quantity at $\abs*{\br{} - \br{}'} = 0$, which ``resembles'' the long-range interaction operator  $\abs*{\br{} - \br{}'}^{-1} \erf(\rsmu{}{} \abs*{\br{} - \br{}'})$ used within RS-DFT. \cite{Giner_2018}
}

\titou{
We start therefore by considering an effective non-divergent two-electron interaction $W^{\Bas}(\br{},\br{}')$ within the basis set which reproduces the expectation value of the Coulomb interaction over a given pair density $\n{2}{\Bas}(\br{},\br{}')$, \ie,
\begin{equation}
\frac{1}{2}	\iint \frac{\n{2}{\Bas}(\br{},\br{}')}{\abs*{\br{} - \br{}'}} d\br{} d\br{}'
	= 
\frac{1}{2}	\iint \n{2}{\Bas}(\br{},\br{}') W^{\Bas}(\br{},\br{}') d\br{} d\br{}'.
\end{equation}
The properties of $W^{\Bas}(\br{},\br{}')$ are detailed in Ref.~\onlinecite{Giner_2018}. A key aspect is that because the value of $W^{\Bas}(\br{},\br{}')$ at coalescence, $W^{\Bas}(\br{},\br{})$, is necessarily finite in a finite basis $\Bas$, one can approximate $W^{\Bas}(\br{},\br{}')$ by a non-divergent, long-range interaction of the form 
\begin{equation}
	\label{eq:Wapprox}
	W^{\Bas}(\br{},\br{}') 
	\approx \frac{1}{2} \qty{ 
		\frac{\erf[\rsmu{}{\Bas}(\br{}) \abs*{\br{} - \br{}'}]}{\abs*{\br{} - \br{}'}}
		+ \frac{\erf[\rsmu{}{\Bas}(\br{}') \abs*{\br{} - \br{}'}]}{\abs*{\br{} - \br{}'}}
		}.
\end{equation}
The information about the finiteness of the basis set is then transferred to the range-separation \textit{function} $\rsmu{}{\Bas}(\br{})$, and its value can be determined by ensuring that the two sides of Eq.~\eqref{eq:Wapprox} are strictly equal at $\abs*{\br{} - \br{}'} = 0$.
Knowing that $\lim_{r \to 0} \erf(\mu r)/r = 2\mu/\sqrt{\pi}$, this yields
\begin{equation}
	\rsmu{}{\Bas}(\br{}) = \frac{\sqrt{\pi}}{2} W^{\Bas}(\br{},\br{}).
\end{equation}
}

\titou{Following Refs.~\onlinecite{Giner_2018,Loos_2019,Giner_2019}, we adopt the following definition for $W^{\Bas}(\br{},\br{}')$ 
\begin{equation}
\label{eq:W}
	W^{\Bas}(\br{},\br{}') =  
	\begin{cases}
		f^{\Bas}(\br{},\br{}')/\n{2}{\Bas}(\br{},\br{}'),		&\text{if  } \n{2}{\Bas}(\br{},\br{}')	\neq 0,		\\
		\infty,									&	\text{otherwise},				\\
	\end{cases}
\end{equation}
where, in this work, $f^{\Bas}(\br{},\br{}')$ and $\n{2}{\Bas}(\br{},\br{}')$ are calculated using the opposite-spin two-electron density matrix of a spin-restricted single determinant (such as HF and KS).
For a closed-shell system, we have
\begin{equation}
\label{fBsum}
	f^{\Bas}(\br{},\br{}') = 2 \sum_{pq}^{\Nbas} \sum_{ij}^{\Nocc} \MO{p}(\br{})\MO{i}(\br{})  (pi|qj)\MO{q}(\br{}')  \MO{j}(\br{}'),
\end{equation}
and 
\begin{eqnarray}
\n{2}{\Bas}(\br{},\br{}') &=& 2  \sum_{ij}^{\Nocc} \MO{i}(\br{})^2  \MO{j}(\br{}')^2
= \frac{1}{2} n^{\Bas}(\br{}) n^{\Bas}(\br{}'),
\end{eqnarray}
where $n^{\Bas}(\br{})$ is the one-electron density. The quantity $\n{2}{\Bas}(\br{},\br{}')$ represents the opposite-spin pair density of a closed-shell system with a single-determinant wave function. Note that in Eq.~\eqref{fBsum} the indices $p$ and $q$ run over all occupied and virtual orbitals ($\Nbas= \Nocc+\Nvirt$ is the total dimension of the basis set).
}

\titou{Thanks to this definition, the effective interaction $W^{\Bas}(\br{},\br{}')$ has the interesting property
\begin{equation}
	\lim_{\Bas \to \CBS}  W^{\Bas}(\br{},\br{}') = \abs*{\br{} - \br{}'}^{-1},
\end{equation}
which means that in the CBS limit one recovers the genuine (divergent) Coulomb interaction. 
Therefore, in the CBS limit, the coalescence value $W^{\Bas}(\br{},\br{})$ goes to infinity, and so does $\rsmu{}{\Bas}(\br{})$. 
Since the present basis-set correction employs complementary short-range correlation potentials from RS-DFT which have the property of going to zero when $\mu$ goes to infinity, the present basis-set correction properly vanishes in the CBS limit. 
%Note also that the divergence condition of $W^{\Bas}(\br{},\br{}')$ in Eq.~\eqref{eq:W} ensures that one-electron systems are free of correction.
}

%%%%%%%%%%%%%%%%%%%%%%%%
\subsection{Short-range correlation functionals}
\label{sec:srDFT}
%%%%%%%%%%%%%%%%%%%%%%%%

The frequency-independent local self-energy $\bSig{}{\Bas}[\n{}{}](\br{},\br{}') = \bpot{}{\Bas}[\n{}{}](\br{}) \delta(\br{}-\br{}')$ originates from the functional derivative of complementary basis-correction density functionals $\bpot{}{\Bas}[\n{}{}](\br{}) = \delta \bE{}{\Bas}[\n{}{}] / \delta \n{}{}(\br{})$. 

\titou{In this work, we have tested two complementary density functionals coming from two approximations to the short-range correlation functional with multideterminant (md) reference of RS-DFT. \cite{Toulouse_2005} 
The first one is a short-range local-density approximation ($\srLDA$) \cite{Toulouse_2005,Paziani_2006}
\begin{equation}
        \label{eq:def_lda_tot}
        \bE{\srLDA}{\Bas}[\n{}{}] =
        \int \n{}{}(\br{}) \be{\text{c,md}}{\srLDA}\qty(\n{}{}(\br{}),\rsmu{}{\Bas}(\br{})) \dbr{},
\end{equation}
where the correlation energy per particle $\be{\text{c,md}}{\srLDA}\qty(\n{}{},\rsmu{}{})$ has been parametrized from calculations on the uniform electron gas \cite{Loos_2016} reported in Ref.~\onlinecite{Paziani_2006}. 
The second one is a short-range Perdew-Burke-Ernzerhof ($\srPBE$) approximation \cite{Ferte_2019, Loos_2019}
\begin{equation}
        \label{eq:def_pbe_tot}
        \bE{\srPBE}{\Bas}[\n{}{}] =
        \int \n{}{}(\br{}) \be{\text{c,md}}{\srPBE}\qty(\n{}{}(\br{}),s(\br{}),\rsmu{}{\Bas}(\br{})) \dbr{},
\end{equation}
where $s(\br{})=\nabla n(\br{})/n(\br{})^{4/3}$ is the reduced density gradient and the correlation energy per particle $\be{\text{c,md}}{\srPBE}\qty(\n{}{},s,\rsmu{}{})$ interpolates between the usual PBE correlation energy per particle \cite{Perdew_1996} at $\mu = 0$ and the exact large-$\mu$ behavior \cite{Toulouse_2004, Gori-Giorgi_2006, Paziani_2006} using the on-top pair density of the Coulombic uniform electron gas (see Ref.~\onlinecite{Loos_2019}). Note that the information on the local basis-set incompleteness error is provided to these RS-DFT functionals through the range-separation function $\rsmu{}{\Bas}(\br{})$.
}

\titou{From these energy functionals, we generate the potentials $\bpot{\srLDA}{\Bas}[\n{}{}](\br{}) = \delta \bE{\srLDA}{\Bas}[\n{}{}]/\delta \n{}{}(\br{})$ and $\bpot{\srPBE}{\Bas}[\n{}{}](\br{}) = \delta \bE{\srPBE}{\Bas}[\n{}{}]/\delta \n{}{}(\br{})$ (considering $\rsmu{}{\Bas}(\br{})$ as being fixed) which are then used to obtain the basis-set corrected {\GOWO} quasiparticle energies
\begin{equation}
	\beGOWO{p} = \eGOWO{p} + \bPot{p}{\Bas},
	\label{eq:QP-corrected}
\end{equation}
with
\begin{equation}
\begin{split}
	\bPot{p}{\Bas}
	& = \int \MO{p}(\br{})  \bpot{}{\Bas}[\n{}{}](\br{}) \MO{p}(\br{}) \dbr{},
\end{split}
\end{equation}
where $\bpot{}{\Bas}[\n{}{}](\br{})=\bpot{\srLDA}{\Bas}[\n{}{}](\br{})$ or $\bpot{\srPBE}{\Bas}[\n{}{}](\br{})$ and the density is calculated from the HF or KS orbitals.
The expressions of these srLDA and srPBE correlation potentials are provided in the {\SI}.}

As evidenced by Eq.~\eqref{eq:QP-corrected}, the present basis-set correction is a non-self-consistent, \textit{post}-{\GW} correction.
Although outside the scope of this study, various other strategies can be potentially designed, for example, within linearized {\GOWO} or self-consistent {\GW} calculations.

%%% TABLE I %%%
\begin{squeezetable}
\begin{table*}
\caption{
IPs (in eV) of the 20 smallest molecules of the GW100 set computed at the {\GOWO}@HF level of theory with various basis sets and corrections.
The mean absolute deviation (MAD), root-mean-square deviation (RMSD), and maximum deviation (MAX) with respect to the {\GOWO}@HF/CBS values are also reported.
\label{tab:GW20_HF}
}
	\begin{ruledtabular}
		\begin{tabular}{lccccccccccccc}
					&	\mc{4}{c}{{\GOWO}@HF}	&	\mc{4}{c}{{\GOWO}@HF+srLDA}		&	\mc{4}{c}{{\GOWO}@HF+srPBE}		&  \mc{1}{c}{{\GOWO}@HF}	\\
						\cline{2-5}					\cline{6-9}							\cline{10-13}						\cline{14-14}
		Mol.		&	cc-pVDZ	&	cc-pVTZ	&	cc-pVQZ	& 	cc-pV5Z	&	cc-pVDZ	&	cc-pVTZ	&	cc-pVQZ	& 	cc-pV5Z	&	cc-pVDZ	&	cc-pVTZ	&	cc-pVQZ	& 	cc-pV5Z		& CBS	\\
		\hline
		\ce{He}		&	24.36	&	24.57	&	24.67	&	24.72	&	24.63	&	24.69	&	24.73	&	24.74	&	24.66	&	24.69	&	24.72	&	24.74	&	24.75	\\
		\ce{Ne}		&	20.87	&	21.39	&	21.63	&	21.73	&	21.38	&	21.67	&	21.80	&	21.84	&	21.56	&	21.73	&	21.81	&	21.83	&	21.82	\\
		\ce{H2}		&	16.25	&	16.48	&	16.56	&	16.58	&	16.42	&	16.54	&	16.58	&	16.60	&	16.42	&	16.53	&	16.58	&	16.60	&	16.61	\\
		\ce{Li2}	&	5.23	&	5.34	&	5.39	&	5.42	&	5.31	&	5.37	&	5.41	&	5.43	&	5.28	&	5.37	&	5.41	&	5.43	&	5.44	\\
		\ce{LiH}	&	7.96	&	8.16	&	8.25	&	8.28	&	8.13	&	8.23	&	8.28	&	8.30	&	8.10	&	8.21	&	8.27	&	8.30	&	8.31	\\
		\ce{HF}		&	15.54	&	16.16	&	16.42	&	16.52	&	16.01	&	16.41	&	16.57	&	16.61	&	16.15	&	16.45	&	16.57	&	16.61	&	16.62	\\
		\ce{Ar}		&	15.40	&	15.72	&	15.93	&	16.08	&	15.85	&	15.98	&	16.09	&	16.18	&	15.91	&	15.99	&	16.08	&	16.17	&	16.15	\\
		\ce{H2O}	&	12.16	&	12.79	&	13.04	&	13.14	&	12.58	&	13.01	&	13.16	&	13.21	&	12.68	&	13.03	&	13.16	&	13.20	&	13.23	\\
		\ce{LiF}	&	10.75	&	11.35	&	11.59	&	11.70	&	11.21	&	11.60	&	11.73	&	11.79	&	11.34	&	11.63	&	11.73	&	11.78	&	11.79	\\
		\ce{HCl}	&	12.40	&	12.77	&	12.96	&	13.05	&	12.79	&	12.99	&	13.10	&	13.13	&	12.83	&	12.99	&	13.09	&	13.12	&	13.12	\\
		\ce{BeO}	&	9.47	&	9.77	&	9.98	&	10.09	&	9.85	&	9.97	&	10.09	&	10.15	&	9.93	&	9.98	&	10.08	&	10.15	&	10.16	\\
		\ce{CO}		&	14.66	&	15.02	&	15.17	&	15.24	&	14.99	&	15.18	&	15.26	&	15.29	&	15.04	&	15.18	&	15.25	&	15.29	&	15.30	\\
		\ce{N2}		&	15.87	&	16.31	&	16.48	&	16.56	&	16.22	&	16.50	&	16.59	&	16.62	&	16.30	&	16.50	&	16.58	&	16.62	&	16.62	\\
		\ce{CH4}	&	14.43	&	14.74	&	14.86	&	14.90	&	14.69	&	14.85	&	14.91	&	14.93	&	14.73	&	14.85	&	14.90	&	14.93	&	14.95	\\
		\ce{BH3}	&	13.35	&	13.64	&	13.74	&	13.78	&	13.57	&	13.73	&	13.78	&	13.80	&	13.58	&	13.72	&	13.78	&	13.80	&	13.82	\\
		\ce{NH3}	&	10.59	&	11.13	&	11.32	&	11.40	&	10.93	&	11.30	&	11.41	&	11.45	&	10.99	&	11.30	&	11.41	&	11.44	&	11.47	\\
		\ce{BF}		&	11.08	&	11.30	&	11.38	&	11.42	&	11.29	&	11.40	&	11.43	&	11.45	&	11.29	&	11.38	&	11.42	&	11.45	&	11.45	\\
		\ce{BN}		&	11.35	&	11.69	&	11.85	&	11.92	&	11.67	&	11.85	&	11.94	&	11.98	&	11.72	&	11.85	&	11.93	&	11.97	&	11.98	\\
		\ce{SH2}	&	10.10	&	10.49	&	10.65	&	10.72	&	10.44	&	10.67	&	10.76	&	10.78	&	10.45	&	10.66	&	10.74	&	10.77	&	10.78	\\
		\ce{F2}		&	15.93	&	16.30	&	16.51	&	16.61	&	16.42	&	16.56	&	16.67	&	16.71	&	16.58	&	16.61	&	16.67	&	16.71	&	16.69	\\
		\hline
		MAD			&	0.66	&	0.30	&	0.13	&	0.06	&	0.33	&	0.13	&	0.04	&	0.01	&	0.27	&	0.12	&	0.04	&	0.01	\\
		RMSD		&	0.71	&	0.32	&	0.14	&	0.06	&	0.37	&	0.14	&	0.04	&	0.01	&	0.30	&	0.13	&	0.05	&	0.01	\\
		MAX			&	1.08	&	0.46	&	0.22	&	0.10	&	0.65	&	0.22	&	0.07	&	0.03	&	0.54	&	0.20	&	0.08	&	0.03	\\
		\end{tabular}
	\end{ruledtabular}
\end{table*}
\end{squeezetable}

%%% TABLE II %%%
\begin{squeezetable}
\begin{table*}
\caption{
IPs (in eV) of the 20 smallest molecules of the GW100 set computed at the {\GOWO}@PBE0 level of theory with various basis sets and corrections.
The mean absolute deviation (MAD), root-mean-square deviation (RMSD), and maximum deviation (MAX) with respect to the {\GOWO}@PBE0/CBS values are also reported.
\label{tab:GW20_PBE0}
}
	\begin{ruledtabular}
		\begin{tabular}{lccccccccccccc}
					&	\mc{4}{c}{{\GOWO}@PBE0}	&	\mc{4}{c}{{\GOWO}@PBE0+srLDA}		&	\mc{4}{c}{{\GOWO}@PBE0+srPBE}	&	\mc{1}{c}{{\GOWO}@PBE0}\\
						\cline{2-5}					\cline{6-9}							\cline{10-13}							\cline{14-14}				
		Mol.		&	cc-pVDZ	&	cc-pVTZ	&	cc-pVQZ	& 	cc-pV5Z	&	cc-pVDZ	&	cc-pVTZ	&	cc-pVQZ	& 	cc-pV5Z	&	cc-pVDZ	&	cc-pVTZ	&	cc-pVQZ	& 	cc-pV5Z	&	CBS		\\
		\hline
		\ce{He}		&	23.99	&	23.98	&	24.03	&	24.04	&	24.26	&	24.09	&	24.09	&	24.07	&	24.29	&	24.10	&	24.08	&	24.07	&	24.06	\\
		\ce{Ne}		&	20.35	&	20.88	&	21.05	&	21.05	&	20.86	&	21.16	&	21.22	&	21.16	&	21.05	&	21.22	&	21.23	&	21.15	&	21.12	\\
		\ce{H2}		&	15.98	&	16.13	&	16.19	&	16.21	&	16.16	&	16.20	&	16.22	&	16.22	&	16.16	&	16.19	&	16.22	&	16.22	&	16.23	\\
		\ce{Li2}	&	5.15	&	5.24	&	5.28	&	5.31	&	5.23	&	5.28	&	5.30	&	5.32	&	5.21	&	5.27	&	5.30	&	5.32	&	5.32	\\
		\ce{LiH}	&	7.32	&	7.49	&	7.56	&	7.59	&	7.48	&	7.55	&	7.59	&	7.61	&	7.45	&	7.54	&	7.58	&	7.61	&	7.62	\\
		\ce{HF}		&	14.95	&	15.61	&	15.82	&	15.85	&	15.41	&	15.85	&	15.97	&	15.94	&	15.56	&	15.89	&	15.97	&	15.93	&	15.94	\\
		\ce{Ar}		&	14.93	&	15.25	&	15.42	&	15.50	&	15.37	&	15.50	&	15.58	&	15.60	&	15.44	&	15.52	&	15.58	&	15.59	&	15.56	\\
		\ce{H2O}	&	11.53	&	12.21	&	12.43	&	12.47	&	11.95	&	12.43	&	12.55	&	12.54	&	12.05	&	12.45	&	12.55	&	12.54	&	12.56	\\
		\ce{LiF}	&	9.89	&	10.60	&	10.82	&	10.94	&	10.35	&	10.84	&	10.96	&	11.02	&	10.48	&	10.87	&	10.96	&	11.02	&	11.02	\\
		\ce{HCl}	&	11.96	&	12.34	&	12.50	&	12.57	&	12.35	&	12.56	&	12.64	&	12.65	&	12.39	&	12.56	&	12.63	&	12.64	&	12.63	\\
		\ce{BeO}	&	9.16	&	9.44	&	9.63	&	9.74	&	9.53	&	9.64	&	9.74	&	9.80	&	9.61	&	9.65	&	9.74	&	9.79	&	9.80	\\
		\ce{CO}		&	13.67	&	14.02	&	14.13	&	14.18	&	14.00	&	14.18	&	14.22	&	14.23	&	14.05	&	14.18	&	14.22	&	14.23	&	14.22	\\
		\ce{N2}		&	14.84	&	15.30	&	15.44	&	15.50	&	15.22	&	15.50	&	15.55	&	15.56	&	15.31	&	15.51	&	15.54	&	15.55	&	15.55	\\
		\ce{CH4}	&	13.85	&	14.15	&	14.27	&	14.30	&	14.11	&	14.27	&	14.32	&	14.33	&	14.15	&	14.27	&	14.32	&	14.33	&	14.35	\\
		\ce{BH3}	&	12.87	&	13.13	&	13.22	&	13.26	&	13.09	&	13.23	&	13.27	&	13.28	&	13.10	&	13.22	&	13.26	&	13.28	&	13.29	\\
		\ce{NH3}	&	9.96	&	10.56	&	10.73	&	10.75	&	10.31	&	10.72	&	10.82	&	10.80	&	10.37	&	10.72	&	10.81	&	10.79	&	10.82	\\
		\ce{BF}		&	10.66	&	10.87	&	10.92	&	10.94	&	10.88	&	10.96	&	10.97	&	10.97	&	10.88	&	10.95	&	10.96	&	10.97	&	10.96	\\
		\ce{BN}		&	11.07	&	11.40	&	11.54	&	11.60	&	11.40	&	11.56	&	11.63	&	11.65	&	11.45	&	11.56	&	11.62	&	11.65	&	11.65	\\
		\ce{SH2}	&	9.69	&	10.10	&	10.25	&	10.30	&	10.03	&	10.28	&	10.35	&	10.36	&	10.04	&	10.27	&	10.34	&	10.35	&	10.36	\\
		\ce{F2}		&	14.92	&	15.38	&	15.57	&	15.64	&	15.41	&	15.65	&	15.73	&	15.74	&	15.57	&	15.69	&	15.73	&	15.73	&	15.71	\\
		\hline
		MAD			&	0.60	&	0.24	&	0.10	&	0.05	&	0.29	&	0.07	&	0.02	&	0.01	&	0.23	&	0.07	&	0.03	&	0.01	\\
		RMSD		&	0.66	&	0.26	&	0.11	&	0.06	&	0.33	&	0.08	&	0.03	&	0.02	&	0.27	&	0.08	&	0.04	&	0.01	\\
		MAX			&	1.12	&	0.42	&	0.19	&	0.09	&	0.67	&	0.18	&	0.09	&	0.04	&	0.54	&	0.15	&	0.10	&	0.03	\\
		\end{tabular}
	\end{ruledtabular}
\end{table*}
\end{squeezetable}

%%%%%%%%%%%%%%%%%%%%%%%%
\section{Computational details}
\label{sec:compdetails}
%%%%%%%%%%%%%%%%%%%%%%%%
All the geometries have been extracted from the GW100 set. \cite{vanSetten_2015}
Unless otherwise stated, all the {\GOWO} calculations have been performed with the MOLGW software developed by Bruneval and coworkers. \cite{Bruneval_2016a}
The HF, PBE, and PBE0 calculations as well as the srLDA and srPBE basis-set corrections have been computed with Quantum Package, \cite{QP2} which by default uses the SG-2 quadrature grid for the numerical integrations.
Frozen-core (FC) calculations are systematically performed. 
The FC density-based basis-set correction~\cite{Loos_2019} is used consistently with the FC approximation in the {\GOWO} calculations.
The {\GOWO} quasiparticle energies have been obtained ``graphically'', \ie, by solving the non-linear, frequency-dependent quasiparticle equation \eqref{eq:QP-G0W0} (without linearization).
Moreover, the infinitesimal $\eta$ in Eq.~\eqref{eq:SigC} has been set to zero.

\titou{Compared to the conventional $\order*{\Nocc^3 \Nvirt^3}$ computational cost of {\GW}, the present basis-set correction represents a marginal $\order*{\Nocc^2 \Nbas^2 \Ngrid}$ additional cost as further discussed in Refs.~\onlinecite{Loos_2019, Giner_2019}.}
Note, however, that the formal $\order*{\Nocc^3 \Nvirt^3}$ computational scaling of {\GW} can be significantly reduced thanks to resolution-of-the-identity techniques \cite{vanSetten_2013, Bruneval_2016, Duchemin_2017} and other tricks. \cite{Rojas_1995, Duchemin_2019}

%%% FIG 1 %%%
\begin{figure*}
	\includegraphics[width=0.45\linewidth]{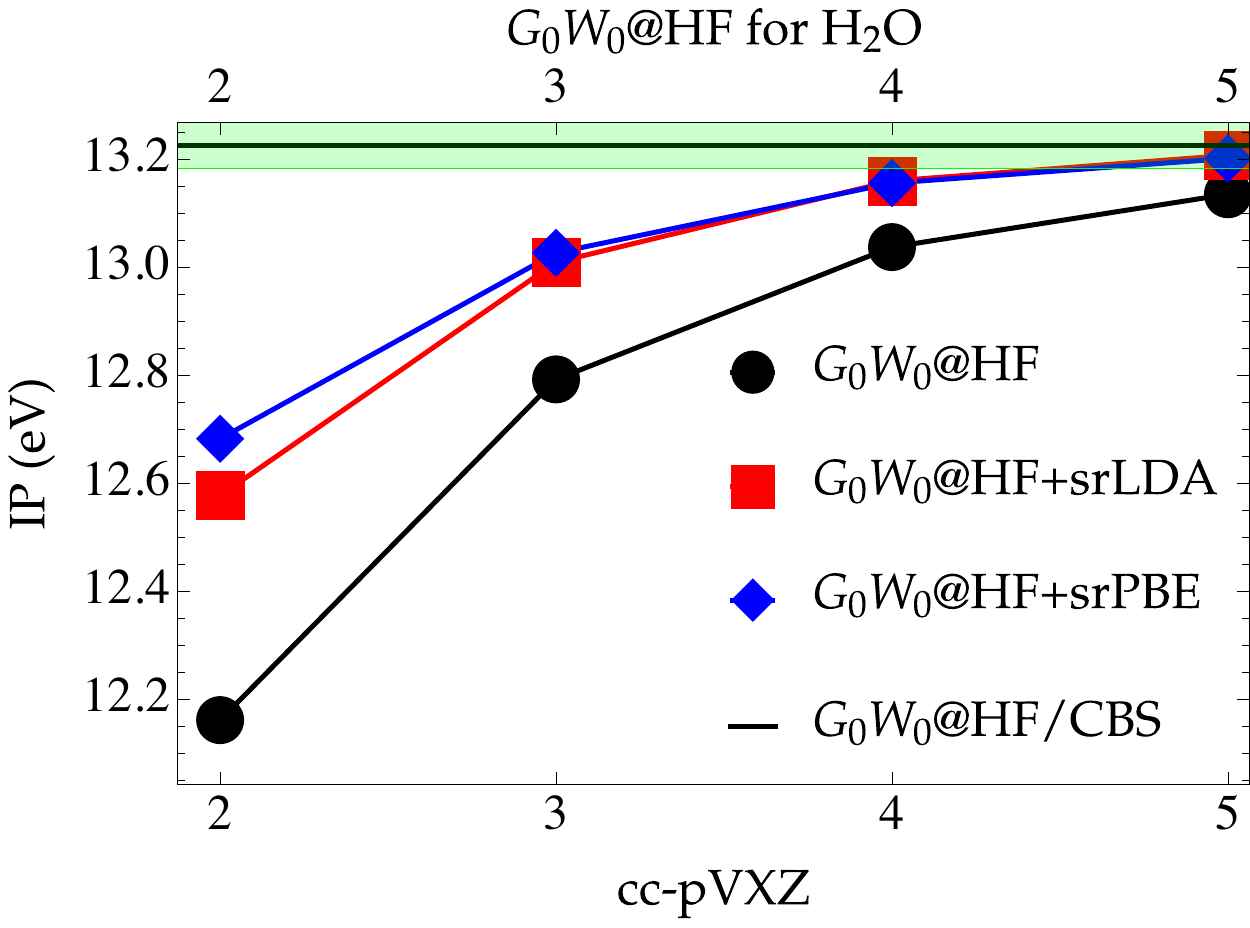}
	\hspace{1cm}
	\includegraphics[width=0.45\linewidth]{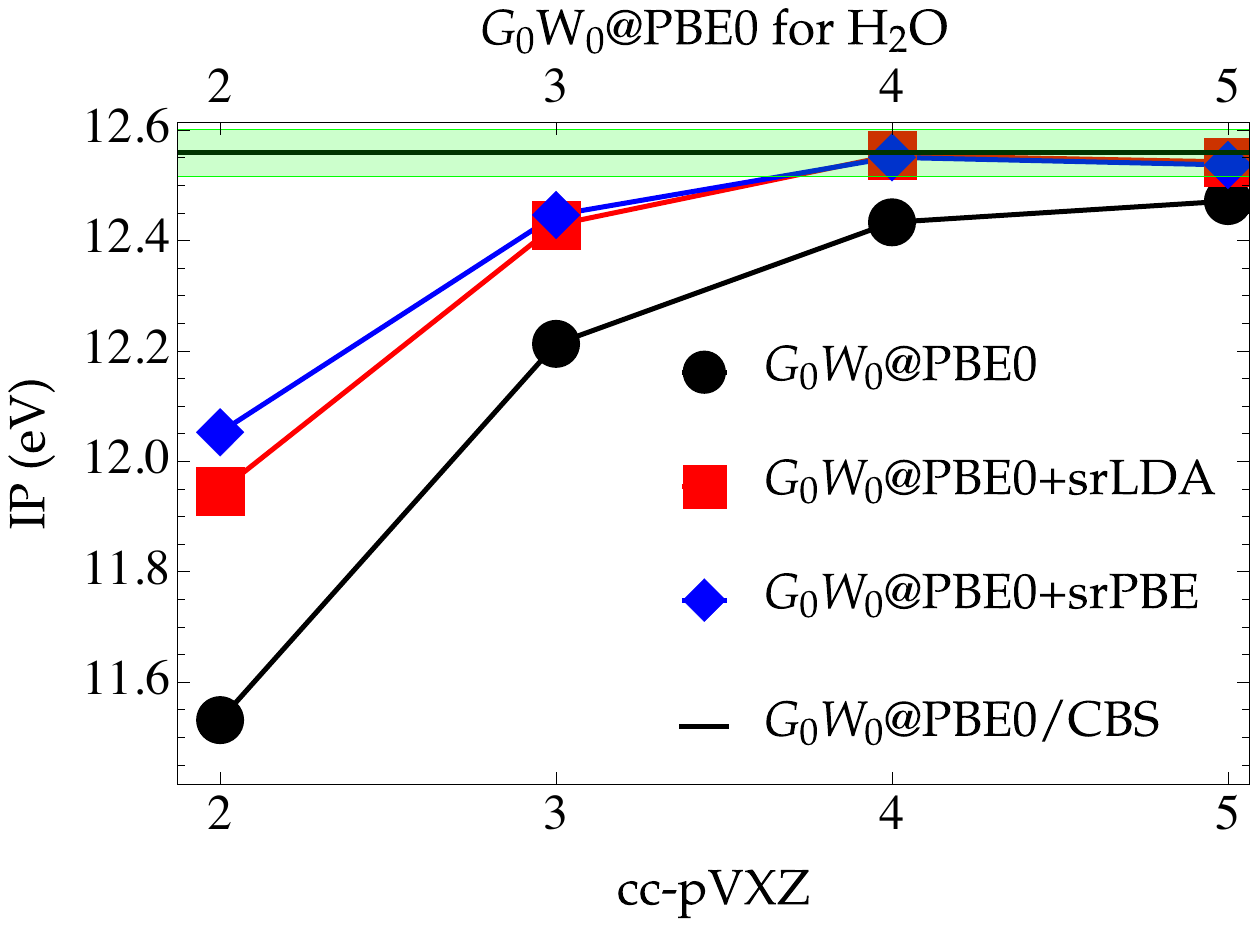}
	\caption{
	IP (in eV) of the water molecule computed at the {\GOWO} (black circles), {\GOWO}+srLDA (red squares), and {\GOWO}+srPBE (blue diamonds) levels of theory with increasingly large Dunning's basis sets \cite{Dunning_1989} (cc-pVDZ, cc-pVTZ, cc-pVQZ, and cc-pV5Z) with two different starting points: HF (left) and PBE0 (right). 
	The thick black line represents the CBS value obtained by extrapolation (see text for more details).
	The green area corresponds to chemical accuracy (\ie, error below $1$ {\kcal} or $0.043$ eV).
	\label{fig:IP_G0W0_H2O}
	}
\end{figure*}

%%%%%%%%%%%%%%%%%%%%%%%%
\section{Results and Discussion}
\label{sec:results}
%%%%%%%%%%%%%%%%%%%%%%%%
In this section, we study a subset of atoms and molecules from the GW100 test set. \cite{vanSetten_2015}
In particular, we study the 20 smallest molecules of the GW100 set, a subset that we label as GW20.
This subset has been recently considered by Lewis and Berkelbach to study the effect of vertex corrections to $\W{}{}$ on IPs of molecules. \cite{Lewis_2019a}
Later in this section, we also study the five canonical nucleobases (adenine, cytosine, thymine, guanine, and uracil) which are also part of the GW100 test set. 

%%%%%%%%%%%%%%%%%%%%%%%%
\subsection{GW20}
\label{sec:GW20}
%%%%%%%%%%%%%%%%%%%%%%%%
The IPs of the GW20 set obtained at the {\GOWO}@{\HF} and {\GOWO}@{\PBEO} levels with increasingly larger Dunning's basis sets cc-pVXZ (X $=$ D, T, Q, and 5) are reported in Tables \ref{tab:GW20_HF} and \ref{tab:GW20_PBE0}, respectively.
The corresponding statistical deviations (with respect to the CBS values) are also reported: mean absolute deviation (MAD), root-mean-square deviation (RMSD), and maximum deviation (MAX). 
These reference CBS values have been obtained with the usual X$^{-3}$ extrapolation procedure using the three largest basis sets. \cite{Bruneval_2012}

The convergence of the IP of the water molecule with respect to the basis set size is depicted in Fig.~\ref{fig:IP_G0W0_H2O}.
This represents a typical example.
Additional graphs reporting the convergence of the IPs of each molecule of the GW20 subset at the {\GOWO}@{\HF} and {\GOWO}@{\PBEO} levels are reported in the {\SI}.

Tables \ref{tab:GW20_HF} and \ref{tab:GW20_PBE0} (as well as Fig.~\ref{fig:IP_G0W0_H2O}) clearly evidence that the present basis-set correction significantly increases the rate of convergence of IPs.
At the {\GOWO}@{\HF} (see Table \ref{tab:GW20_HF}), the MAD of the conventional calculations (\textit{i.e}, without basis-set correction) is roughly divided by two each time one increases the basis set size (MADs of $0.60$, $0.24$, $0.10$, and $0.05$ eV going from cc-pVDZ to cc-pV5Z) with maximum errors higher than $1$ eV for molecules such as \ce{HF}, \ce{H2O}, and \ce{LiF} with the smallest basis set.
Even with the largest quintuple-$\zeta$ basis, the MAD is still above chemical accuracy (\ie, error below $1$ {\kcal} or $0.043$ eV).

For each basis set, the correction brought by the short-range correlation functionals reduces by roughly half or more the MAD, RMSD, and MAX compared to the correction-free calculations.
For example, we obtain MADs of $0.27$, $0.12$, $0.04$, and $0.01$ eV at the {\GOWO}@HF+srPBE level with increasingly larger basis sets.
Interestingly, in most cases, the srPBE correction is slightly larger than the srLDA one.
This observation is clear at the cc-pVDZ level but, for larger basis sets, the two RS-DFT-based corrections are essentially equivalent.
Note also that, in some cases, the corrected IPs slightly overshoot the CBS values. 
However, it is hard to know if it is not due to the extrapolation error.
In a nutshell, the present basis-set correction provides cc-pVQZ quality results at the cc-pVTZ level.
Besides, it allows to reach chemical accuracy with the quadruple-$\zeta$ basis set, an accuracy that could not be reached even with the cc-pV5Z basis set for the conventional calculations.

Very similar conclusions are drawn at the {\GOWO}@{\PBEO} level (see Table \ref{tab:GW20_PBE0}) with a slightly faster convergence to the CBS limit.
For example, at the {\GOWO}@PBE0+srLDA/cc-pVQZ level, the MAD is only $0.02$ eV with a maximum error as small as $0.09$ eV. 

It is worth pointing out that, for ground-state properties such as atomization and correlation energies, the density-based correction brought a larger acceleration of the basis-set convergence.
For example, we evidenced in Ref.~\onlinecite{Loos_2019} that quintuple-$\zeta$ quality atomization and correlation energies are recovered with triple-$\zeta$ basis sets.
Here, the overall gain seems to be less important.
The possible reasons for this could be: i) DFT approximations are usually less accurate for the potential than for the energy, \cite{Kim_2013} and ii) because the present scheme only corrects the basis-set incompleteness error originating from the electron-electron cusp, some incompleteness remains at the HF or KS level. \cite{Adler_2007}

%%% TABLE III %%%
\begin{table*}
\caption{
IPs (in eV) of the five canonical nucleobases (adenine, cytosine, thymine, guanine, and uracil) computed at the {\GOWO}@PBE level of theory for various basis sets and corrections.
The deviation with respect to the {\GOWO}@PBE/def2-TQZVP extrapolated values are reported in square brackets. 
The extrapolation error is reported in parenthesis.
\titou{Extrapolated {\GOWO}@PBE results obtained with plane-wave basis sets, as well as CCSD(T) and experimental results are reported for comparison.}
\label{tab:DNA_IP}
}
	\begin{ruledtabular}
		\begin{tabular}{llccccc}
									&				&	\mc{5}{c}{IPs of nucleobases (eV)}					\\	
														\cline{3-7}										
		Method						&	Basis		&	\tabc{Adenine} 	&	\tabc{Cytosine}	&	\tabc{Guanine}		&	\tabc{Thymine}	& \tabc{Uracil}	\\	
		\hline
		{\GOWO}@PBE\fnm[1]			&	def2-SVP	&	$7.27$[$-0.88$]	&	$7.53$[$-0.92$]	&	$6.95$[$-0.92$]	&	$8.02$[$-0.85$]	&	$8.38$[$-1.00$]	\\
		{\GOWO}@PBE+srLDA\fnm[1]	&	def2-SVP	&	$7.60$[$-0.55$]	&	$7.95$[$-0.50$]	&	$7.29$[$-0.59$]	&	$8.36$[$-0.51$]	&	$8.80$[$-0.58$]	\\
		{\GOWO}@PBE+srPBE\fnm[1]	&	def2-SVP	&	$7.64$[$-0.51$]	&	$8.06$[$-0.39$]	&	$7.34$[$-0.54$]	&	$8.41$[$-0.45$]	&	$8.91$[$-0.47$]	\\
		{\GOWO}@PBE\fnm[1]			&	def2-TZVP	&	$7.74$[$-0.41$]	&	$8.06$[$-0.39$]	&	$7.45$[$-0.42$]	&	$8.48$[$-0.38$]	&	$8.86$[$-0.52$]	\\
		{\GOWO}@PBE+srLDA\fnm[1]	&	def2-TZVP	&	$7.92$[$-0.23$]	&	$8.26$[$-0.19$]	&	$7.64$[$-0.23$]	&	$8.67$[$-0.20$]	&	$9.25$[$-0.13$]	\\
		{\GOWO}@PBE+srPBE\fnm[1]	&	def2-TZVP	&	$7.92$[$-0.23$]	&	$8.27$[$-0.18$]	&	$7.64$[$-0.23$]	&	$8.68$[$-0.19$]	&	$9.27$[$-0.11$]	\\
		{\GOWO}@PBE\fnm[2]			&	def2-QZVP	&	$7.98$[$-0.18$]	&	$8.29$[$-0.16$]	&	$7.69$[$-0.18$]	&	$8.71$[$-0.16$]	&	$9.22$[$-0.16$]	\\
		{\GOWO}@PBE\fnm[3]			&	def2-TQZVP	&	$8.16(1)$		&	$8.44(1)$		&	$7.87(1)$		&	$8.87(1)$		&	$9.38(1)$		\\
		\hline
		{\GOWO}@PBE\fnm[4]			&	plane waves	&	$8.12$			&	$8.40$			&	$7.85$			&	$8.83$			&	$9.36$			\\
		{\GOWO}@PBE\fnm[5]			&	plane waves	&	$8.09(2)$		&	$8.40(2)$		&	$7.82(2)$		&	$8.82(2)$		&	$9.19(2)$		\\
		\hline
		CCSD(T)\fnm[6]				&	aug-cc-pVDZ	&	$8.40$			&	$8.76$			&	$8.09$			&	$9.04$			&	$9.43$			\\
		CCSD(T)\fnm[7]				&	def2-TZVPP	&	$8.33$			&	$9.51$			&	$8.03$			&	$9.08$			&	$10.13$			\\
		\hline
		Experiment\fnm[8]			&				&	$8.48$			&	$8.94$			&	$8.24$			&	$9.20$			&	$9.68$			\\
		\end{tabular}
	\end{ruledtabular}
	\fnt[1]{This work.}
	\fnt[2]{Unpublished data taken from \url{https://gw100.wordpress.com} obtained with TURBOMOLE v7.0.}
	\fnt[3]{Extrapolated values obtained from the def2-TZVP and def2-QZVP values.}
	\fnt[4]{\titou{Extrapolated plane-wave results from Ref.~\onlinecite{Maggio_2017} obtained with WEST.}}
	\fnt[5]{\titou{Extrapolated plane-wave results from Ref.~\onlinecite{Govoni_2018} obtained with VASP.}}
	\fnt[6]{\titou{CCSD(T)//CCSD/aug-cc-pVDZ results from Ref.~\onlinecite{Roca-Sanjuan_2006}.}}
	\fnt[7]{Reference \onlinecite{Krause_2015}.}
	\fnt[8]{Experimental values are taken from Ref.~\onlinecite{vanSetten_2015} and correspond to vertical ionization energies.}
\end{table*}

%%% FIG 2 %%%
\begin{figure*}
	\includegraphics[width=\linewidth]{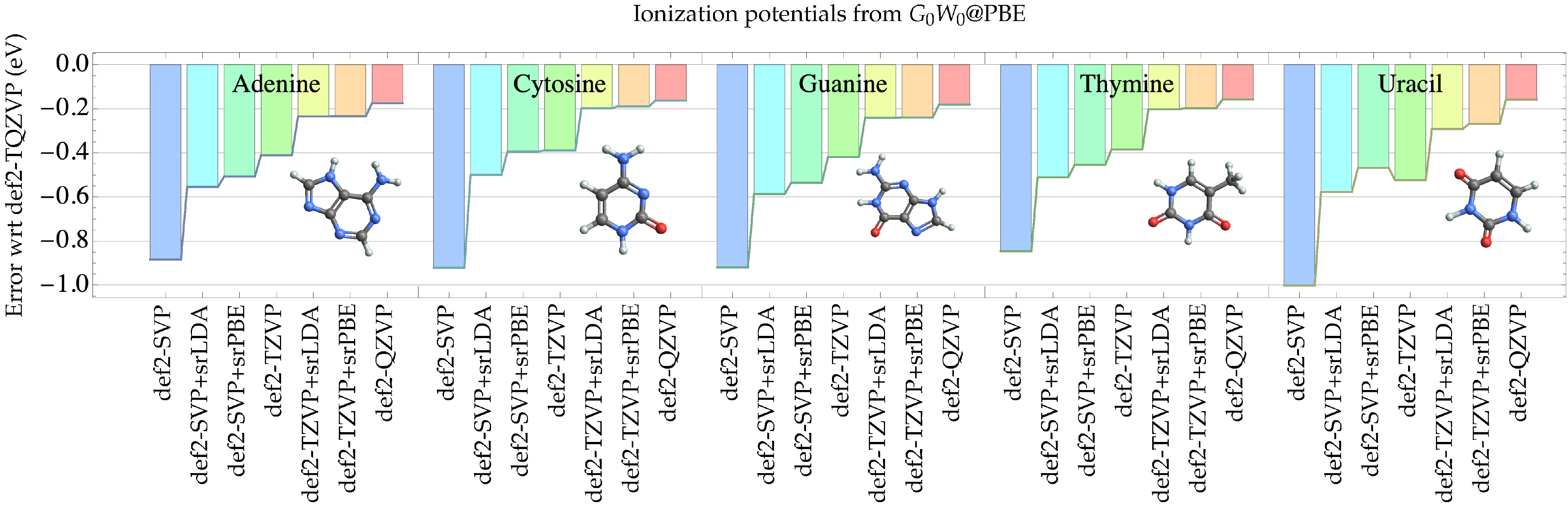}
	\caption{
	Error (in eV) with respect to the {\GOWO}@PBE/def2-TQZVP extrapolated values for the IPs of the five canonical nucleobases (adenine, cytosine, thymine, guanine, and uracil) computed at the {\GOWO}@PBE level of theory for various basis sets and corrections.
	\label{fig:DNA_IP}
	}
\end{figure*}

%%%%%%%%%%%%%%%%%%%%%%%%
\subsection{Nucleobases}
\label{sec:DNA}
%%%%%%%%%%%%%%%%%%%%%%%%
In order to check the transferability of the present observations to larger systems, we have computed the values of the IPs of the five canonical nucleobases (adenine, cytosine, thymine, guanine, and uracil) at the {\GOWO}@PBE level of theory with a different basis set family. \cite{Weigend_2003a, Weigend_2005a}
The numerical values are reported in Table \ref{tab:DNA_IP}, and their error with respect to the {\GOWO}@PBE/def2-TQZVP extrapolated values \cite{vanSetten_2015} (obtained via extrapolation of the def2-TZVP and def2-QZVP results) are shown in Fig.~\ref{fig:DNA_IP}.
\titou{Table \ref{tab:DNA_IP} also contains extrapolated IPs obtained with plane-wave basis sets with two different software packages. \cite{Maggio_2017,Govoni_2018}
The CCSD(T)/def2-TZVPP computed by Krause \textit{et al.} \cite{Krause_2015} on the same geometries, the CCSD(T)//CCSD/aug-cc-pVDZ results from Ref.~\onlinecite{Roca-Sanjuan_2006}, as well as the experimental results extracted from Ref.~\onlinecite{vanSetten_2015} are reported for comparison purposes.}

For these five systems, the IPs are all of the order of $8$ or $9$ eV with an amplitude of roughly $1$ eV between the smallest basis set (def2-SVP) and the CBS value.
The conclusions that we have drawn in the previous subsection do apply here as well.
For the smallest double-$\zeta$ basis def2-SVP, the basis-set correction reduces by roughly half an eV the basis-set incompleteness error.
It is particularly interesting to note that the basis-set corrected def2-TZVP results are on par with the correction-free def2-QZVP numbers.
This is quite remarkable as the number of basis functions jumps from $371$ to $777$ for the largest system (guanine).

%%%%%%%%%%%%%%%%%%%%%%%%
\section{Conclusion}
\label{sec:conclusion}
%%%%%%%%%%%%%%%%%%%%%%%%
In the present manuscript, we have shown that the density-based basis-set correction developed by some of the authors in Ref.~\onlinecite{Giner_2018} and applied recently to ground- and excited-state properties \cite{Loos_2019, Giner_2019} can also be successfully applied to Green function methods such as {\GW}.
In particular, we have evidenced that the present basis-set correction (which relies on LDA- or PBE-based short-range correlation functionals) significantly speeds up the convergence of IPs for small and larger molecules towards the CBS limit.
These findings have been observed for different {\GW} starting points (HF, PBE, and PBE0).
\titou{We have observed that the performance of the two short-range correlation functionals (srLDA and srPBE) are quite similar with a slight edge for srPBE over srLDA.
Therefore, because srPBE is only slightly more computationally expensive than srLDA, we do recommend the use of srPBE.}

As mentioned earlier, the present basis-set correction can be straightforwardly applied to other properties of interest such as electron affinities or fundamental gaps.
It is also applicable to other flavors of {\GW} such as the partially self-consistent {\evGW} or {\qsGW} methods, \titou{and more generally to any approximation of the self-energy.}
We are currently investigating the performance of the present approach within linear response theory in order to speed up the convergence of excitation energies obtained within the RPA and Bethe-Salpeter equation (BSE) \cite{Strinati_1988, Leng_2016, Blase_2018} formalisms.  
We hope to report on this in the near future.

%%%%%%%%%%%%%%%%%%%%%%%%
\section*{Supporting Information}
%%%%%%%%%%%%%%%%%%%%%%%%
See {\SI} for \titou{the expression of the short-range correlation potentials}, additional graphs reporting the convergence of the ionization potentials of the GW20 subset with respect to the size of the basis set, \titou{and
the numerical data of Tables \ref{tab:GW20_HF} and \ref{tab:GW20_PBE0} (provided in txt and json formats).}

%%%%%%%%%%%%%%%%%%%%%%%%
\begin{acknowledgements}
PFL would like to thank Fabien Bruneval for technical assistance. 
PFL and JT would like to thank Arjan Berger and Pina Romaniello for stimulating discussions.
This work was performed using HPC resources from GENCI-TGCC (Grant No.~2018-A0040801738) and CALMIP (Toulouse) under allocation 2019-18005.
Funding from the \textit{``Centre National de la Recherche Scientifique''} is acknowledged.
This work has been supported through the EUR grant NanoX ANR-17-EURE-0009 in the framework of the \textit{``Programme des Investissements d'Avenir''}.
\end{acknowledgements}
%%%%%%%%%%%%%%%%%%%%%%%%

\bibliography{GW-srDFT,GW-srDFT-control}

\providecommand{\latin}[1]{#1}
\makeatletter
\providecommand{\doi}
  {\begingroup\let\do\@makeother\dospecials
  \catcode`\{=1 \catcode`\}=2 \doi@aux}
\providecommand{\doi@aux}[1]{\endgroup\texttt{#1}}
\makeatother
\providecommand*\mcitethebibliography{\thebibliography}
\csname @ifundefined\endcsname{endmcitethebibliography}
  {\let\endmcitethebibliography\endthebibliography}{}
\begin{mcitethebibliography}{113}
\providecommand*\natexlab[1]{#1}
\providecommand*\mciteSetBstSublistMode[1]{}
\providecommand*\mciteSetBstMaxWidthForm[2]{}
\providecommand*\mciteBstWouldAddEndPuncttrue
  {\def\EndOfBibitem{\unskip.}}
\providecommand*\mciteBstWouldAddEndPunctfalse
  {\let\EndOfBibitem\relax}
\providecommand*\mciteSetBstMidEndSepPunct[3]{}
\providecommand*\mciteSetBstSublistLabelBeginEnd[3]{}
\providecommand*\EndOfBibitem{}
\mciteSetBstSublistMode{f}
\mciteSetBstMaxWidthForm{subitem}{(\alph{mcitesubitemcount})}
\mciteSetBstSublistLabelBeginEnd
  {\mcitemaxwidthsubitemform\space}
  {\relax}
  {\relax}

\bibitem[Martin \latin{et~al.}(2016)Martin, Reining, and Ceperley]{Martin_2016}
Martin,~R.~M.; Reining,~L.; Ceperley,~D.~M. \emph{Interacting Electrons: Theory
  and Computational Approaches}; Cambridge University Press, 2016\relax
\mciteBstWouldAddEndPuncttrue
\mciteSetBstMidEndSepPunct{\mcitedefaultmidpunct}
{\mcitedefaultendpunct}{\mcitedefaultseppunct}\relax
\EndOfBibitem
\bibitem[Aryasetiawan and Gunnarsson(1998)Aryasetiawan, and
  Gunnarsson]{Aryasetiawan_1998}
Aryasetiawan,~F.; Gunnarsson,~O. The GW method. \emph{Rep. Prog. Phys.}
  \textbf{1998}, \emph{61}, 237--312\relax
\mciteBstWouldAddEndPuncttrue
\mciteSetBstMidEndSepPunct{\mcitedefaultmidpunct}
{\mcitedefaultendpunct}{\mcitedefaultseppunct}\relax
\EndOfBibitem
\bibitem[Onida \latin{et~al.}(2002)Onida, Reining, and and]{Onida_2002}
Onida,~G.; Reining,~L.; and,~A.~R. Electronic excitations: density-functional
  versus many-body Green's-function approaches. \emph{Rev. Mod. Phys.}
  \textbf{2002}, \emph{74}, 601--659\relax
\mciteBstWouldAddEndPuncttrue
\mciteSetBstMidEndSepPunct{\mcitedefaultmidpunct}
{\mcitedefaultendpunct}{\mcitedefaultseppunct}\relax
\EndOfBibitem
\bibitem[Reining(2017)]{Reining_2017}
Reining,~L. The {{GW}} Approximation: Content, Successes and Limitations: {{The
  GW}} Approximation. \emph{Wiley Interdiscip. Rev. Comput. Mol. Sci.}
  \textbf{2017}, e1344\relax
\mciteBstWouldAddEndPuncttrue
\mciteSetBstMidEndSepPunct{\mcitedefaultmidpunct}
{\mcitedefaultendpunct}{\mcitedefaultseppunct}\relax
\EndOfBibitem
\bibitem[Blase \latin{et~al.}(2011)Blase, Attaccalite, and Olevano]{Blase_2011}
Blase,~X.; Attaccalite,~C.; Olevano,~V. First-Principles {{GW}} Calculations
  for Fullerenes, Porphyrins, Phtalocyanine, and Other Molecules of Interest
  for Organic Photovoltaic Applications. \emph{Phys. Rev. B} \textbf{2011},
  \emph{83}, 115103\relax
\mciteBstWouldAddEndPuncttrue
\mciteSetBstMidEndSepPunct{\mcitedefaultmidpunct}
{\mcitedefaultendpunct}{\mcitedefaultseppunct}\relax
\EndOfBibitem
\bibitem[Faber \latin{et~al.}(2011)Faber, Attaccalite, Olevano, Runge, and
  Blase]{Faber_2011}
Faber,~C.; Attaccalite,~C.; Olevano,~V.; Runge,~E.; Blase,~X. First-Principles
  {{GW}} Calculations for {{DNA}} and {{RNA}} Nucleobases. \emph{Phys. Rev. B}
  \textbf{2011}, \emph{83}, 115123\relax
\mciteBstWouldAddEndPuncttrue
\mciteSetBstMidEndSepPunct{\mcitedefaultmidpunct}
{\mcitedefaultendpunct}{\mcitedefaultseppunct}\relax
\EndOfBibitem
\bibitem[Bruneval(2012)]{Bruneval_2012}
Bruneval,~F. Ionization Energy of Atoms Obtained from {{{\emph{GW}}}}
  Self-Energy or from Random Phase Approximation Total Energies. \emph{J. Chem.
  Phys.} \textbf{2012}, \emph{136}, 194107\relax
\mciteBstWouldAddEndPuncttrue
\mciteSetBstMidEndSepPunct{\mcitedefaultmidpunct}
{\mcitedefaultendpunct}{\mcitedefaultseppunct}\relax
\EndOfBibitem
\bibitem[Bruneval \latin{et~al.}(2015)Bruneval, Hamed, and
  Neaton]{Bruneval_2015}
Bruneval,~F.; Hamed,~S.~M.; Neaton,~J.~B. A Systematic Benchmark of the
  {\emph{Ab Initio}} {{Bethe}}-{{Salpeter}} Equation Approach for Low-Lying
  Optical Excitations of Small Organic Molecules. \emph{J. Chem. Phys.}
  \textbf{2015}, \emph{142}, 244101\relax
\mciteBstWouldAddEndPuncttrue
\mciteSetBstMidEndSepPunct{\mcitedefaultmidpunct}
{\mcitedefaultendpunct}{\mcitedefaultseppunct}\relax
\EndOfBibitem
\bibitem[Bruneval \latin{et~al.}(2016)Bruneval, Rangel, Hamed, Shao, Yang, and
  Neaton]{Bruneval_2016}
Bruneval,~F.; Rangel,~T.; Hamed,~S.~M.; Shao,~M.; Yang,~C.; Neaton,~J.~B. Molgw
  1: {{Many}}-Body Perturbation Theory Software for Atoms, Molecules, and
  Clusters. \emph{Comput. Phys. Commun.} \textbf{2016}, \emph{208},
  149--161\relax
\mciteBstWouldAddEndPuncttrue
\mciteSetBstMidEndSepPunct{\mcitedefaultmidpunct}
{\mcitedefaultendpunct}{\mcitedefaultseppunct}\relax
\EndOfBibitem
\bibitem[Bruneval(2016)]{Bruneval_2016a}
Bruneval,~F. Optimized Virtual Orbital Subspace for Faster {{GW}} Calculations
  in Localized Basis. \emph{J. Chem. Phys.} \textbf{2016}, \emph{145},
  234110\relax
\mciteBstWouldAddEndPuncttrue
\mciteSetBstMidEndSepPunct{\mcitedefaultmidpunct}
{\mcitedefaultendpunct}{\mcitedefaultseppunct}\relax
\EndOfBibitem
\bibitem[Boulanger \latin{et~al.}(2014)Boulanger, Jacquemin, Duchemin, and
  Blase]{Boulanger_2014}
Boulanger,~P.; Jacquemin,~D.; Duchemin,~I.; Blase,~X. Fast and {{Accurate
  Electronic Excitations}} in {{Cyanines}} with the {{Many}}-{{Body
  Bethe}}\textendash{}{{Salpeter Approach}}. \emph{J. Chem. Theory Comput.}
  \textbf{2014}, \emph{10}, 1212--1218\relax
\mciteBstWouldAddEndPuncttrue
\mciteSetBstMidEndSepPunct{\mcitedefaultmidpunct}
{\mcitedefaultendpunct}{\mcitedefaultseppunct}\relax
\EndOfBibitem
\bibitem[Blase \latin{et~al.}(2016)Blase, Boulanger, Bruneval, Fernandez-Serra,
  and Duchemin]{Blase_2016}
Blase,~X.; Boulanger,~P.; Bruneval,~F.; Fernandez-Serra,~M.; Duchemin,~I.
  {{{\emph{GW}}}} and {{Bethe}}-{{Salpeter}} Study of Small Water Clusters.
  \emph{J. Chem. Phys.} \textbf{2016}, \emph{144}, 034109\relax
\mciteBstWouldAddEndPuncttrue
\mciteSetBstMidEndSepPunct{\mcitedefaultmidpunct}
{\mcitedefaultendpunct}{\mcitedefaultseppunct}\relax
\EndOfBibitem
\bibitem[Li \latin{et~al.}(2017)Li, Holzmann, Duchemin, Blase, and
  Olevano]{Li_2017}
Li,~J.; Holzmann,~M.; Duchemin,~I.; Blase,~X.; Olevano,~V. Helium {{Atom
  Excitations}} by the {{G W}} and {{Bethe}}-{{Salpeter Many}}-{{Body
  Formalism}}. \emph{Phys. Rev. Lett.} \textbf{2017}, \emph{118}, 163001\relax
\mciteBstWouldAddEndPuncttrue
\mciteSetBstMidEndSepPunct{\mcitedefaultmidpunct}
{\mcitedefaultendpunct}{\mcitedefaultseppunct}\relax
\EndOfBibitem
\bibitem[Hung \latin{et~al.}(2016)Hung, {da Jornada}, Souto-Casares,
  Chelikowsky, Louie, and {\"O}{\u g}{\"u}t]{Hung_2016}
Hung,~L.; {da Jornada},~F.~H.; Souto-Casares,~J.; Chelikowsky,~J.~R.;
  Louie,~S.~G.; {\"O}{\u g}{\"u}t,~S. Excitation Spectra of Aromatic Molecules
  within a Real-Space {{G W}} -{{BSE}} Formalism: {{Role}} of Self-Consistency
  and Vertex Corrections. \emph{Phys. Rev. B} \textbf{2016}, \emph{94},
  085125\relax
\mciteBstWouldAddEndPuncttrue
\mciteSetBstMidEndSepPunct{\mcitedefaultmidpunct}
{\mcitedefaultendpunct}{\mcitedefaultseppunct}\relax
\EndOfBibitem
\bibitem[Hung \latin{et~al.}(2017)Hung, Bruneval, Baishya, and {\"O}{\u
  g}{\"u}t]{Hung_2017}
Hung,~L.; Bruneval,~F.; Baishya,~K.; {\"O}{\u g}{\"u}t,~S. Benchmarking the
  {{{\emph{GW}}}} {{Approximation}} and {{Bethe}}\textendash{}{{Salpeter
  Equation}} for {{Groups IB}} and {{IIB Atoms}} and {{Monoxides}}. \emph{J.
  Chem. Theory Comput.} \textbf{2017}, \emph{13}, 2135--2146\relax
\mciteBstWouldAddEndPuncttrue
\mciteSetBstMidEndSepPunct{\mcitedefaultmidpunct}
{\mcitedefaultendpunct}{\mcitedefaultseppunct}\relax
\EndOfBibitem
\bibitem[{van Setten} \latin{et~al.}(2015){van Setten}, Caruso, Sharifzadeh,
  Ren, Scheffler, Liu, Lischner, Lin, Deslippe, Louie, Yang, Weigend, Neaton,
  Evers, and Rinke]{vanSetten_2015}
{van Setten},~M.~J.; Caruso,~F.; Sharifzadeh,~S.; Ren,~X.; Scheffler,~M.;
  Liu,~F.; Lischner,~J.; Lin,~L.; Deslippe,~J.~R.; Louie,~S.~G.; Yang,~C.;
  Weigend,~F.; Neaton,~J.~B.; Evers,~F.; Rinke,~P. {{{\emph{GW}}}} 100:
  {{Benchmarking}} {{{\emph{G}}}} {\textsubscript{0}} {{{\emph{W}}}}
  {\textsubscript{0}} for {{Molecular Systems}}. \emph{J. Chem. Theory Comput.}
  \textbf{2015}, \emph{11}, 5665--5687\relax
\mciteBstWouldAddEndPuncttrue
\mciteSetBstMidEndSepPunct{\mcitedefaultmidpunct}
{\mcitedefaultendpunct}{\mcitedefaultseppunct}\relax
\EndOfBibitem
\bibitem[{van Setten} \latin{et~al.}(2018){van Setten}, Costa, Vi{\~n}es, and
  Illas]{vanSetten_2018}
{van Setten},~M.~J.; Costa,~R.; Vi{\~n}es,~F.; Illas,~F. Assessing
  {{{\emph{GW}}}} {{Approaches}} for {{Predicting Core Level Binding
  Energies}}. \emph{J. Chem. Theory Comput.} \textbf{2018}, \emph{14},
  877--883\relax
\mciteBstWouldAddEndPuncttrue
\mciteSetBstMidEndSepPunct{\mcitedefaultmidpunct}
{\mcitedefaultendpunct}{\mcitedefaultseppunct}\relax
\EndOfBibitem
\bibitem[Ou and Subotnik(2016)Ou, and Subotnik]{Ou_2016}
Ou,~Q.; Subotnik,~J.~E. Comparison between {{{\emph{GW}}}} and
  {{Wave}}-{{Function}}-{{Based Approaches}}: {{Calculating}} the {{Ionization
  Potential}} and {{Electron Affinity}} for {{1D Hubbard Chains}}. \emph{J.
  Phys. Chem. A} \textbf{2016}, \emph{120}, 4514--4525\relax
\mciteBstWouldAddEndPuncttrue
\mciteSetBstMidEndSepPunct{\mcitedefaultmidpunct}
{\mcitedefaultendpunct}{\mcitedefaultseppunct}\relax
\EndOfBibitem
\bibitem[Ou and Subotnik(2018)Ou, and Subotnik]{Ou_2018}
Ou,~Q.; Subotnik,~J.~E. Comparison between the {{Bethe}}\textendash{}{{Salpeter
  Equation}} and {{Configuration Interaction Approaches}} for {{Solving}} a
  {{Quantum Chemistry Problem}}: {{Calculating}} the {{Excitation Energy}} for
  {{Finite 1D Hubbard Chains}}. \emph{J. Chem. Theory Comput.} \textbf{2018},
  \emph{14}, 527--542\relax
\mciteBstWouldAddEndPuncttrue
\mciteSetBstMidEndSepPunct{\mcitedefaultmidpunct}
{\mcitedefaultendpunct}{\mcitedefaultseppunct}\relax
\EndOfBibitem
\bibitem[Faber(2014)]{Faber_2014}
Faber,~C. Electronic, Excitonic and Polaronic Properties of Organic Systems
  within the Many-Body {{GW}} and {{Bethe}}-{{Salpeter}} Formalisms: Towards
  Organic Photovoltaics. {{PhD Thesis}}, Universit{\'e} de Grenoble, 2014\relax
\mciteBstWouldAddEndPuncttrue
\mciteSetBstMidEndSepPunct{\mcitedefaultmidpunct}
{\mcitedefaultendpunct}{\mcitedefaultseppunct}\relax
\EndOfBibitem
\bibitem[Marini \latin{et~al.}(2009)Marini, Hogan, Gruning, and
  Varsano]{Marini_2009}
Marini,~A.; Hogan,~C.; Gruning,~M.; Varsano,~D. Yambo: An Ab Initio Tool For
  Excited State Calculations. \emph{Comp. Phys. Comm.} \textbf{2009},
  \emph{180}, 1392\relax
\mciteBstWouldAddEndPuncttrue
\mciteSetBstMidEndSepPunct{\mcitedefaultmidpunct}
{\mcitedefaultendpunct}{\mcitedefaultseppunct}\relax
\EndOfBibitem
\bibitem[Deslippe \latin{et~al.}(2012)Deslippe, Samsonidze, Strubbe, Jain,
  Cohen, and Louie]{Deslippe_2012}
Deslippe,~J.; Samsonidze,~G.; Strubbe,~D.~A.; Jain,~M.; Cohen,~M.~L.;
  Louie,~S.~G. BerkeleyGW: A Massively Parallel Computer Package for the
  Calculation of the Quasiparticle and Optical Properties of Materials and
  Nanostructures. \emph{Comput. Phys. Commun.} \textbf{2012}, \emph{183},
  1269\relax
\mciteBstWouldAddEndPuncttrue
\mciteSetBstMidEndSepPunct{\mcitedefaultmidpunct}
{\mcitedefaultendpunct}{\mcitedefaultseppunct}\relax
\EndOfBibitem
\bibitem[Maggio \latin{et~al.}(2017)Maggio, Liu, {van Setten}, and
  Kresse]{Maggio_2017}
Maggio,~E.; Liu,~P.; {van Setten},~M.~J.; Kresse,~G. {{{\emph{GW}}}} 100: {{A
  Plane Wave Perspective}} for {{Small Molecules}}. \emph{J. Chem. Theory
  Comput.} \textbf{2017}, \emph{13}, 635--648\relax
\mciteBstWouldAddEndPuncttrue
\mciteSetBstMidEndSepPunct{\mcitedefaultmidpunct}
{\mcitedefaultendpunct}{\mcitedefaultseppunct}\relax
\EndOfBibitem
\bibitem[Blase \latin{et~al.}(2018)Blase, Duchemin, and Jacquemin]{Blase_2018}
Blase,~X.; Duchemin,~I.; Jacquemin,~D. The {{Bethe}}\textendash{}{{Salpeter}}
  Equation in Chemistry: Relations with {{TD}}-{{DFT}}, Applications and
  Challenges. \emph{Chem. Soc. Rev.} \textbf{2018}, \emph{47}, 1022--1043\relax
\mciteBstWouldAddEndPuncttrue
\mciteSetBstMidEndSepPunct{\mcitedefaultmidpunct}
{\mcitedefaultendpunct}{\mcitedefaultseppunct}\relax
\EndOfBibitem
\bibitem[{van Setten} \latin{et~al.}(2013){van Setten}, Weigend, and
  Evers]{vanSetten_2013}
{van Setten},~M.~J.; Weigend,~F.; Evers,~F. The {{{\emph{GW}}}} -{{Method}} for
  {{Quantum Chemistry Applications}}: {{Theory}} and {{Implementation}}.
  \emph{J. Chem. Theory Comput.} \textbf{2013}, \emph{9}, 232--246\relax
\mciteBstWouldAddEndPuncttrue
\mciteSetBstMidEndSepPunct{\mcitedefaultmidpunct}
{\mcitedefaultendpunct}{\mcitedefaultseppunct}\relax
\EndOfBibitem
\bibitem[Kaplan \latin{et~al.}(2015)Kaplan, Weigend, Evers, and {van
  Setten}]{Kaplan_2015}
Kaplan,~F.; Weigend,~F.; Evers,~F.; {van Setten},~M.~J. Off-{{Diagonal
  Self}}-{{Energy Terms}} and {{Partially Self}}-{{Consistency}} in
  {{{\emph{GW}}}} {{Calculations}} for {{Single Molecules}}: {{Efficient
  Implementation}} and {{Quantitative Effects}} on {{Ionization Potentials}}.
  \emph{J. Chem. Theory Comput.} \textbf{2015}, \emph{11}, 5152--5160\relax
\mciteBstWouldAddEndPuncttrue
\mciteSetBstMidEndSepPunct{\mcitedefaultmidpunct}
{\mcitedefaultendpunct}{\mcitedefaultseppunct}\relax
\EndOfBibitem
\bibitem[Kaplan \latin{et~al.}(2016)Kaplan, Harding, Seiler, Weigend, Evers,
  and {van Setten}]{Kaplan_2016}
Kaplan,~F.; Harding,~M.~E.; Seiler,~C.; Weigend,~F.; Evers,~F.; {van
  Setten},~M.~J. Quasi-{{Particle Self}}-{{Consistent}} {{{\emph{GW}}}} for
  {{Molecules}}. \emph{J. Chem. Theory Comput.} \textbf{2016}, \emph{12},
  2528--2541\relax
\mciteBstWouldAddEndPuncttrue
\mciteSetBstMidEndSepPunct{\mcitedefaultmidpunct}
{\mcitedefaultendpunct}{\mcitedefaultseppunct}\relax
\EndOfBibitem
\bibitem[Krause and Klopper(2017)Krause, and Klopper]{Krause_2017}
Krause,~K.; Klopper,~W. Implementation Of The {{Bethe}}-{{Salpeter}} Equation
  In The {{Turbomole}} Program. \emph{J. Comput. Chem.} \textbf{2017},
  \emph{38}, 383--388\relax
\mciteBstWouldAddEndPuncttrue
\mciteSetBstMidEndSepPunct{\mcitedefaultmidpunct}
{\mcitedefaultendpunct}{\mcitedefaultseppunct}\relax
\EndOfBibitem
\bibitem[Caruso \latin{et~al.}(2012)Caruso, Rinke, Ren, Scheffler, and
  Rubio]{Caruso_2012}
Caruso,~F.; Rinke,~P.; Ren,~X.; Scheffler,~M.; Rubio,~A. Unified Description of
  Ground and Excited States of Finite Systems: {{The}} Self-Consistent {{G W}}
  Approach. \emph{Phys. Rev. B} \textbf{2012}, \emph{86}, 081102(R)\relax
\mciteBstWouldAddEndPuncttrue
\mciteSetBstMidEndSepPunct{\mcitedefaultmidpunct}
{\mcitedefaultendpunct}{\mcitedefaultseppunct}\relax
\EndOfBibitem
\bibitem[Caruso \latin{et~al.}(2013)Caruso, Rohr, Hellgren, Ren, Rinke, Rubio,
  and Scheffler]{Caruso_2013}
Caruso,~F.; Rohr,~D.~R.; Hellgren,~M.; Ren,~X.; Rinke,~P.; Rubio,~A.;
  Scheffler,~M. Bond {{Breaking}} and {{Bond Formation}}: {{How Electron
  Correlation}} Is {{Captured}} in {{Many}}-{{Body Perturbation Theory}} and
  {{Density}}-{{Functional Theory}}. \emph{Phys. Rev. Lett.} \textbf{2013},
  \emph{110}, 146403\relax
\mciteBstWouldAddEndPuncttrue
\mciteSetBstMidEndSepPunct{\mcitedefaultmidpunct}
{\mcitedefaultendpunct}{\mcitedefaultseppunct}\relax
\EndOfBibitem
\bibitem[Caruso \latin{et~al.}(2013)Caruso, Rinke, Ren, Rubio, and
  Scheffler]{Caruso_2013a}
Caruso,~F.; Rinke,~P.; Ren,~X.; Rubio,~A.; Scheffler,~M. Self-Consistent {{G
  W}} : {{All}}-Electron Implementation with Localized Basis Functions.
  \emph{Phys. Rev. B} \textbf{2013}, \emph{88}, 075105\relax
\mciteBstWouldAddEndPuncttrue
\mciteSetBstMidEndSepPunct{\mcitedefaultmidpunct}
{\mcitedefaultendpunct}{\mcitedefaultseppunct}\relax
\EndOfBibitem
\bibitem[Caruso(2013)]{Caruso_2013b}
Caruso,~F. Self-Consistent {{GW}} Approach for the Unified Description of
  Ground and Excited States of Finite Systems. {{PhD Thesis}}, Freie
  Universit{\"a}t Berlin, 2013\relax
\mciteBstWouldAddEndPuncttrue
\mciteSetBstMidEndSepPunct{\mcitedefaultmidpunct}
{\mcitedefaultendpunct}{\mcitedefaultseppunct}\relax
\EndOfBibitem
\bibitem[Hedin(1965)]{Hedin_1965}
Hedin,~L. New Method for Calculating the One-Particle {{Green}}'s Function with
  Application to the Electron-Gas Problem. \emph{Phys. Rev.} \textbf{1965},
  \emph{139}, A796\relax
\mciteBstWouldAddEndPuncttrue
\mciteSetBstMidEndSepPunct{\mcitedefaultmidpunct}
{\mcitedefaultendpunct}{\mcitedefaultseppunct}\relax
\EndOfBibitem
\bibitem[Olver \latin{et~al.}(2010)Olver, Lozier, Boisvert, and
  Clark]{NISTbook}
Olver,~F. W.~J., Lozier,~D.~W., Boisvert,~R.~F., Clark,~C.~W., Eds. \emph{NIST
  Handbook of Mathematical Functions}; Cambridge University Press: New York,
  2010\relax
\mciteBstWouldAddEndPuncttrue
\mciteSetBstMidEndSepPunct{\mcitedefaultmidpunct}
{\mcitedefaultendpunct}{\mcitedefaultseppunct}\relax
\EndOfBibitem
\bibitem[Loos \latin{et~al.}(2018)Loos, Romaniello, and Berger]{Loos_2018}
Loos,~P.~F.; Romaniello,~P.; Berger,~J.~A. Green functions and
  self-consistency: insights from the spherium model. \emph{J. Chem. Theory
  Comput.} \textbf{2018}, \emph{14}, 3071--3082\relax
\mciteBstWouldAddEndPuncttrue
\mciteSetBstMidEndSepPunct{\mcitedefaultmidpunct}
{\mcitedefaultendpunct}{\mcitedefaultseppunct}\relax
\EndOfBibitem
\bibitem[Hybertsen and Louie(1985)Hybertsen, and Louie]{Hybertsen_1985a}
Hybertsen,~M.~S.; Louie,~S.~G. First-{{Principles Theory}} of
  {{Quasiparticles}}: {{Calculation}} of {{Band Gaps}} in {{Semiconductors}}
  and {{Insulators}}. \emph{Phys. Rev. Lett.} \textbf{1985}, \emph{55},
  1418--1421\relax
\mciteBstWouldAddEndPuncttrue
\mciteSetBstMidEndSepPunct{\mcitedefaultmidpunct}
{\mcitedefaultendpunct}{\mcitedefaultseppunct}\relax
\EndOfBibitem
\bibitem[Hybertsen and Louie(1986)Hybertsen, and Louie]{Hybertsen_1986}
Hybertsen,~M.~S.; Louie,~S.~G. Electron Correlation in Semiconductors and
  Insulators: {{Band}} Gaps and Quasiparticle Energies. \emph{Phys. Rev. B}
  \textbf{1986}, \emph{34}, 5390--5413\relax
\mciteBstWouldAddEndPuncttrue
\mciteSetBstMidEndSepPunct{\mcitedefaultmidpunct}
{\mcitedefaultendpunct}{\mcitedefaultseppunct}\relax
\EndOfBibitem
\bibitem[Bruneval and Marques(2013)Bruneval, and Marques]{Bruneval_2013}
Bruneval,~F.; Marques,~M. A.~L. Benchmarking the {{Starting Points}} of the
  {{{\emph{GW}}}} {{Approximation}} for {{Molecules}}. \emph{J. Chem. Theory
  Comput.} \textbf{2013}, \emph{9}, 324--329\relax
\mciteBstWouldAddEndPuncttrue
\mciteSetBstMidEndSepPunct{\mcitedefaultmidpunct}
{\mcitedefaultendpunct}{\mcitedefaultseppunct}\relax
\EndOfBibitem
\bibitem[Jacquemin \latin{et~al.}(2016)Jacquemin, Duchemin, and
  Blase]{Jacquemin_2016}
Jacquemin,~D.; Duchemin,~I.; Blase,~X. Assessment Of The Convergence Of
  Partially Self-Consistent {{BSE/GW}} Calculations. \emph{Mol. Phys.}
  \textbf{2016}, \emph{114}, 957\relax
\mciteBstWouldAddEndPuncttrue
\mciteSetBstMidEndSepPunct{\mcitedefaultmidpunct}
{\mcitedefaultendpunct}{\mcitedefaultseppunct}\relax
\EndOfBibitem
\bibitem[Gui \latin{et~al.}(2018)Gui, Holzer, and Klopper]{Gui_2018}
Gui,~X.; Holzer,~C.; Klopper,~W. Accuracy {{Assessment}} of {{{\emph{GW}}}}
  {{Starting Points}} for {{Calculating Molecular Excitation Energies Using}}
  the {{Bethe}}\textendash{{Salpeter Formalism}}. \emph{J. Chem. Theory
  Comput.} \textbf{2018}, \emph{14}, 2127--2136\relax
\mciteBstWouldAddEndPuncttrue
\mciteSetBstMidEndSepPunct{\mcitedefaultmidpunct}
{\mcitedefaultendpunct}{\mcitedefaultseppunct}\relax
\EndOfBibitem
\bibitem[Shishkin and Kresse(2007)Shishkin, and Kresse]{Shishkin_2007}
Shishkin,~M.; Kresse,~G. Self-Consistent {{G W}} Calculations for
  Semiconductors and Insulators. \emph{Phys. Rev. B} \textbf{2007}, \emph{75},
  235102\relax
\mciteBstWouldAddEndPuncttrue
\mciteSetBstMidEndSepPunct{\mcitedefaultmidpunct}
{\mcitedefaultendpunct}{\mcitedefaultseppunct}\relax
\EndOfBibitem
\bibitem[Ke(2011)]{Ke_2011}
Ke,~S.-H. All-Electron {{G W}} Methods Implemented in Molecular Orbital Space:
  {{Ionization}} Energy and Electron Affinity of Conjugated Molecules.
  \emph{Phys. Rev. B} \textbf{2011}, \emph{84}, 205415\relax
\mciteBstWouldAddEndPuncttrue
\mciteSetBstMidEndSepPunct{\mcitedefaultmidpunct}
{\mcitedefaultendpunct}{\mcitedefaultseppunct}\relax
\EndOfBibitem
\bibitem[Koval \latin{et~al.}(2014)Koval, Foerster, and
  S{\'a}nchez-Portal]{Koval_2014}
Koval,~P.; Foerster,~D.; S{\'a}nchez-Portal,~D. Fully Self-Consistent {{G W}}
  and Quasiparticle Self-Consistent {{G W}} for Molecules. \emph{Phys. Rev. B}
  \textbf{2014}, \emph{89}, 155417\relax
\mciteBstWouldAddEndPuncttrue
\mciteSetBstMidEndSepPunct{\mcitedefaultmidpunct}
{\mcitedefaultendpunct}{\mcitedefaultseppunct}\relax
\EndOfBibitem
\bibitem[Jacquemin \latin{et~al.}(2017)Jacquemin, Duchemin, Blondel, and
  Blase]{Jacquemin_2017}
Jacquemin,~D.; Duchemin,~I.; Blondel,~A.; Blase,~X. Benchmark of
  {{Bethe}}-{{Salpeter}} for {{Triplet Excited}}-{{States}}. \emph{J. Chem.
  Theory Comput.} \textbf{2017}, \emph{13}, 767--783\relax
\mciteBstWouldAddEndPuncttrue
\mciteSetBstMidEndSepPunct{\mcitedefaultmidpunct}
{\mcitedefaultendpunct}{\mcitedefaultseppunct}\relax
\EndOfBibitem
\bibitem[Kato(1957)]{Kato_1957}
Kato,~T. On The Eigenfunctions Of Many-Particle Systems In Quantum Mechanics.
  \emph{Commun. Pure Appl. Math.} \textbf{1957}, \emph{10}, 151\relax
\mciteBstWouldAddEndPuncttrue
\mciteSetBstMidEndSepPunct{\mcitedefaultmidpunct}
{\mcitedefaultendpunct}{\mcitedefaultseppunct}\relax
\EndOfBibitem
\bibitem[Hylleraas(1929)]{Hylleraas_1929}
Hylleraas,~E.~A. Neue Berechnung der Energie des Heliums im Grundzustande,
  sowie des tiefsten Terms von Ortho-Helium. \emph{Z. Phys.} \textbf{1929},
  \emph{54}, 347\relax
\mciteBstWouldAddEndPuncttrue
\mciteSetBstMidEndSepPunct{\mcitedefaultmidpunct}
{\mcitedefaultendpunct}{\mcitedefaultseppunct}\relax
\EndOfBibitem
\bibitem[Kutzelnigg(1985)]{Kutzelnigg_1985}
Kutzelnigg,~W. R12-Dependent Terms In The Wave Function As Closed Sums Of
  Partial Wave Amplitudes For Large L. \emph{Theor. Chim. Acta} \textbf{1985},
  \emph{68}, 445\relax
\mciteBstWouldAddEndPuncttrue
\mciteSetBstMidEndSepPunct{\mcitedefaultmidpunct}
{\mcitedefaultendpunct}{\mcitedefaultseppunct}\relax
\EndOfBibitem
\bibitem[Noga and Kutzelnigg(1994)Noga, and Kutzelnigg]{Noga_1994}
Noga,~J.; Kutzelnigg,~W. Coupled Cluster Theory That Takes Care Of The
  Correlation Cusp By Inclusion Of Linear Terms In The Interelectronic
  Coordinates. \emph{J. Chem. Phys.} \textbf{1994}, \emph{101}, 7738\relax
\mciteBstWouldAddEndPuncttrue
\mciteSetBstMidEndSepPunct{\mcitedefaultmidpunct}
{\mcitedefaultendpunct}{\mcitedefaultseppunct}\relax
\EndOfBibitem
\bibitem[Kutzelnigg and Klopper(1991)Kutzelnigg, and Klopper]{Kutzelnigg_1991}
Kutzelnigg,~W.; Klopper,~W. Wave Functions With Terms Linear In The
  Interelectronic Coordinates To Take Care Of The Correlation Cusp. I. General
  Theory. \emph{J. Chem. Phys.} \textbf{1991}, \emph{94}, 1985\relax
\mciteBstWouldAddEndPuncttrue
\mciteSetBstMidEndSepPunct{\mcitedefaultmidpunct}
{\mcitedefaultendpunct}{\mcitedefaultseppunct}\relax
\EndOfBibitem
\bibitem[Kong \latin{et~al.}(2012)Kong, Bischo, and Valeev]{Kong_2012}
Kong,~L.; Bischo,~F.~A.; Valeev,~E.~F. Explicitly Correlated R12/F12 Methods
  for Electronic Structure. \emph{Chem. Rev.} \textbf{2012}, \emph{112},
  75\relax
\mciteBstWouldAddEndPuncttrue
\mciteSetBstMidEndSepPunct{\mcitedefaultmidpunct}
{\mcitedefaultendpunct}{\mcitedefaultseppunct}\relax
\EndOfBibitem
\bibitem[Hattig \latin{et~al.}(2012)Hattig, Klopper, Kohn, and
  Tew]{Hattig_2012}
Hattig,~C.; Klopper,~W.; Kohn,~A.; Tew,~D.~P. Explicitly Correlated Electrons
  in Molecules. \emph{Chem. Rev.} \textbf{2012}, \emph{112}, 4\relax
\mciteBstWouldAddEndPuncttrue
\mciteSetBstMidEndSepPunct{\mcitedefaultmidpunct}
{\mcitedefaultendpunct}{\mcitedefaultseppunct}\relax
\EndOfBibitem
\bibitem[Ten-no and Noga(2012)Ten-no, and Noga]{Tenno_2012a}
Ten-no,~S.; Noga,~J. Explicitly Correlated Electronic Structure Theory From
  R12/F12 Ansatze. \emph{WIREs Comput. Mol. Sci.} \textbf{2012}, \emph{2},
  114\relax
\mciteBstWouldAddEndPuncttrue
\mciteSetBstMidEndSepPunct{\mcitedefaultmidpunct}
{\mcitedefaultendpunct}{\mcitedefaultseppunct}\relax
\EndOfBibitem
\bibitem[Ten-no(2012)]{Tenno_2012b}
Ten-no,~S. Explicitly Correlated Wave Functions: Summary And Perspective.
  \emph{Theor. Chem. Acc.} \textbf{2012}, \emph{131}, 1070\relax
\mciteBstWouldAddEndPuncttrue
\mciteSetBstMidEndSepPunct{\mcitedefaultmidpunct}
{\mcitedefaultendpunct}{\mcitedefaultseppunct}\relax
\EndOfBibitem
\bibitem[Gr\"uneis \latin{et~al.}(2017)Gr\"uneis, Hirata, Ohnishi, and
  Ten-no]{Gruneis_2017}
Gr\"uneis,~A.; Hirata,~S.; Ohnishi,~Y.-Y.; Ten-no,~S. Perspective: Explicitly
  Correlated Electronic Structure Theory For Complex Systems. \emph{J. Chem.
  Phys.} \textbf{2017}, \emph{146}, 080901\relax
\mciteBstWouldAddEndPuncttrue
\mciteSetBstMidEndSepPunct{\mcitedefaultmidpunct}
{\mcitedefaultendpunct}{\mcitedefaultseppunct}\relax
\EndOfBibitem
\bibitem[Tew \latin{et~al.}(2007)Tew, Klopper, Neiss, and Hattig]{Tew_2007}
Tew,~D.~P.; Klopper,~W.; Neiss,~C.; Hattig,~C. Quintuple-{{$\zeta$}} Quality
  Coupled-Cluster Correlation Energies With Triple-{{$\zeta$}} Basis Sets.
  \emph{Phys. Chem. Chem. Phys.} \textbf{2007}, \emph{9}, 1921\relax
\mciteBstWouldAddEndPuncttrue
\mciteSetBstMidEndSepPunct{\mcitedefaultmidpunct}
{\mcitedefaultendpunct}{\mcitedefaultseppunct}\relax
\EndOfBibitem
\bibitem[Giner \latin{et~al.}(2018)Giner, Pradines, Fert\'e, Assaraf, Savin,
  and Toulouse]{Giner_2018}
Giner,~E.; Pradines,~B.; Fert\'e,~A.; Assaraf,~R.; Savin,~A.; Toulouse,~J.
  Curing Basis-Set Convergence Of Wave-Function Theory Using Density-Functional
  Theory: A Systematically Improvable Approach. \emph{J. Chem. Phys.}
  \textbf{2018}, \emph{149}, 194301\relax
\mciteBstWouldAddEndPuncttrue
\mciteSetBstMidEndSepPunct{\mcitedefaultmidpunct}
{\mcitedefaultendpunct}{\mcitedefaultseppunct}\relax
\EndOfBibitem
\bibitem[Loos \latin{et~al.}(2019)Loos, Pradines, Scemama, Toulouse, and
  Giner]{Loos_2019}
Loos,~P.~F.; Pradines,~B.; Scemama,~A.; Toulouse,~J.; Giner,~E. A Density-Based
  Basis-Set Correction for Wave Function Theory. \emph{J. Phys. Chem. Lett.}
  \textbf{2019}, \emph{10}, 2931--2937\relax
\mciteBstWouldAddEndPuncttrue
\mciteSetBstMidEndSepPunct{\mcitedefaultmidpunct}
{\mcitedefaultendpunct}{\mcitedefaultseppunct}\relax
\EndOfBibitem
\bibitem[Giner \latin{et~al.}(2019)Giner, Scemama, Toulouse, and
  Loos]{Giner_2019}
Giner,~E.; Scemama,~A.; Toulouse,~J.; Loos,~P.~F. Chemically Accurate
  Excitation Energies With Small Basis Sets. \emph{J. Chem. Phys.}
  \textbf{2019}, \emph{151}, 144118\relax
\mciteBstWouldAddEndPuncttrue
\mciteSetBstMidEndSepPunct{\mcitedefaultmidpunct}
{\mcitedefaultendpunct}{\mcitedefaultseppunct}\relax
\EndOfBibitem
\bibitem[Barca \latin{et~al.}(2016)Barca, Loos, and Gill]{Barca_2016}
Barca,~G. M.~J.; Loos,~P.-F.; Gill,~P. M.~W. Many-{{Electron Integrals}} over
  {{Gaussian Basis Functions}}. {{I}}. {{Recurrence Relations}} for
  {{Three}}-{{Electron Integrals}}. \emph{Journal of Chemical Theory and
  Computation} \textbf{2016}, \emph{12}, 1735--1740\relax
\mciteBstWouldAddEndPuncttrue
\mciteSetBstMidEndSepPunct{\mcitedefaultmidpunct}
{\mcitedefaultendpunct}{\mcitedefaultseppunct}\relax
\EndOfBibitem
\bibitem[Barca and Loos(2017)Barca, and Loos]{Barca_2017}
Barca,~G.~M.; Loos,~P.-F. Three-and Four-Electron Integrals Involving
  {{Gaussian}} Geminals: {{Fundamental}} Integrals, Upper Bounds, and
  Recurrence Relations. \emph{The Journal of chemical physics} \textbf{2017},
  \emph{147}, 024103\relax
\mciteBstWouldAddEndPuncttrue
\mciteSetBstMidEndSepPunct{\mcitedefaultmidpunct}
{\mcitedefaultendpunct}{\mcitedefaultseppunct}\relax
\EndOfBibitem
\bibitem[Barca and Loos(2018)Barca, and Loos]{Barca_2018}
Barca,~G.~M.; Loos,~P.-F. Recurrence {{Relations}} for {{Four}}-{{Electron
  Integrals Over Gaussian Basis Functions}}. In \emph{Advances in {{Quantum
  Chemistry}}}; {Elsevier}, 2018; Vol.~76; pp 147--165\relax
\mciteBstWouldAddEndPuncttrue
\mciteSetBstMidEndSepPunct{\mcitedefaultmidpunct}
{\mcitedefaultendpunct}{\mcitedefaultseppunct}\relax
\EndOfBibitem
\bibitem[Szabo and Ostlund(1989)Szabo, and Ostlund]{SzaboBook}
Szabo,~A.; Ostlund,~N.~S. \emph{Modern quantum chemistry}; McGraw-Hill: New
  York, 1989\relax
\mciteBstWouldAddEndPuncttrue
\mciteSetBstMidEndSepPunct{\mcitedefaultmidpunct}
{\mcitedefaultendpunct}{\mcitedefaultseppunct}\relax
\EndOfBibitem
\bibitem[Casida and Chong(1989)Casida, and Chong]{Casida_1989}
Casida,~M.~E.; Chong,~D.~P. Physical Interpretation and Assessment of the
  {{Coulomb}}-Hole and Screened-Exchange Approximation for Molecules.
  \emph{Phys. Rev. A} \textbf{1989}, \emph{40}, 4837--4848\relax
\mciteBstWouldAddEndPuncttrue
\mciteSetBstMidEndSepPunct{\mcitedefaultmidpunct}
{\mcitedefaultendpunct}{\mcitedefaultseppunct}\relax
\EndOfBibitem
\bibitem[Casida and Chong(1991)Casida, and Chong]{Casida_1991}
Casida,~M.~E.; Chong,~D.~P. Simplified {{Green}}-Function Approximations:
  {{Further}} Assessment of a Polarization Model for Second-Order Calculation
  of Outer-Valence Ionization Potentials in Molecules. \emph{Phys. Rev. A}
  \textbf{1991}, \emph{44}, 5773--5783\relax
\mciteBstWouldAddEndPuncttrue
\mciteSetBstMidEndSepPunct{\mcitedefaultmidpunct}
{\mcitedefaultendpunct}{\mcitedefaultseppunct}\relax
\EndOfBibitem
\bibitem[Stefanucci and van Leeuwen(2013)Stefanucci, and van
  Leeuwen]{Stefanucci_2013}
Stefanucci,~G.; van Leeuwen,~R. \emph{Nonequilibrium Many-Body Theory of
  Quantum Systems: A Modern Introduction}; {Cambridge University Press}:
  Cambridge, 2013\relax
\mciteBstWouldAddEndPuncttrue
\mciteSetBstMidEndSepPunct{\mcitedefaultmidpunct}
{\mcitedefaultendpunct}{\mcitedefaultseppunct}\relax
\EndOfBibitem
\bibitem[Ortiz(2013)]{Ortiz_2013}
Ortiz,~J.~V. Electron Propagator Theory: An Approach to Prediction and
  Interpretation in Quantum Chemistry: {{Electron}} Propagator Theory.
  \emph{Wiley Interdiscip. Rev. Comput. Mol. Sci.} \textbf{2013}, \emph{3},
  123--142\relax
\mciteBstWouldAddEndPuncttrue
\mciteSetBstMidEndSepPunct{\mcitedefaultmidpunct}
{\mcitedefaultendpunct}{\mcitedefaultseppunct}\relax
\EndOfBibitem
\bibitem[Phillips and Zgid(2014)Phillips, and Zgid]{Phillips_2014}
Phillips,~J.~J.; Zgid,~D. Communication: {{The}} Description of Strong
  Correlation within Self-Consistent {{Green}}'s Function Second-Order
  Perturbation Theory. \emph{J. Chem. Phys.} \textbf{2014}, \emph{140},
  241101\relax
\mciteBstWouldAddEndPuncttrue
\mciteSetBstMidEndSepPunct{\mcitedefaultmidpunct}
{\mcitedefaultendpunct}{\mcitedefaultseppunct}\relax
\EndOfBibitem
\bibitem[Phillips \latin{et~al.}(2015)Phillips, Kananenka, and
  Zgid]{Phillips_2015}
Phillips,~J.~J.; Kananenka,~A.~A.; Zgid,~D. Fractional Charge and Spin Errors
  in Self-Consistent {{Green}}'s Function Theory. \emph{J. Chem. Phys.}
  \textbf{2015}, \emph{142}, 194108\relax
\mciteBstWouldAddEndPuncttrue
\mciteSetBstMidEndSepPunct{\mcitedefaultmidpunct}
{\mcitedefaultendpunct}{\mcitedefaultseppunct}\relax
\EndOfBibitem
\bibitem[Rusakov \latin{et~al.}(2014)Rusakov, Phillips, and Zgid]{Rusakov_2014}
Rusakov,~A.~A.; Phillips,~J.~J.; Zgid,~D. Local {{Hamiltonians}} for
  Quantitative {{Green}}'s Function Embedding Methods. \emph{J. Chem. Phys.}
  \textbf{2014}, \emph{141}, 194105\relax
\mciteBstWouldAddEndPuncttrue
\mciteSetBstMidEndSepPunct{\mcitedefaultmidpunct}
{\mcitedefaultendpunct}{\mcitedefaultseppunct}\relax
\EndOfBibitem
\bibitem[Rusakov and Zgid(2016)Rusakov, and Zgid]{Rusakov_2016}
Rusakov,~A.~A.; Zgid,~D. Self-Consistent Second-Order {{Green}}'s Function
  Perturbation Theory for Periodic Systems. \emph{J. Chem. Phys.}
  \textbf{2016}, \emph{144}, 054106\relax
\mciteBstWouldAddEndPuncttrue
\mciteSetBstMidEndSepPunct{\mcitedefaultmidpunct}
{\mcitedefaultendpunct}{\mcitedefaultseppunct}\relax
\EndOfBibitem
\bibitem[Hirata \latin{et~al.}(2015)Hirata, Hermes, Simons, and
  Ortiz]{Hirata_2015}
Hirata,~S.; Hermes,~M.~R.; Simons,~J.; Ortiz,~J.~V. General-{{Order
  Many}}-{{Body Green}}'s {{Function Method}}. \emph{J. Chem. Theory Comput.}
  \textbf{2015}, \emph{11}, 1595--1606\relax
\mciteBstWouldAddEndPuncttrue
\mciteSetBstMidEndSepPunct{\mcitedefaultmidpunct}
{\mcitedefaultendpunct}{\mcitedefaultseppunct}\relax
\EndOfBibitem
\bibitem[Hirata \latin{et~al.}(2017)Hirata, Doran, Knowles, and
  Ortiz]{Hirata_2017}
Hirata,~S.; Doran,~A.~E.; Knowles,~P.~J.; Ortiz,~J.~V. One-Particle Many-Body
  {{Green}}'s Function Theory: {{Algebraic}} Recursive Definitions,
  Linked-Diagram Theorem, Irreducible-Diagram Theorem, and General-Order
  Algorithms. \emph{J. Chem. Phys.} \textbf{2017}, \emph{147}, 044108\relax
\mciteBstWouldAddEndPuncttrue
\mciteSetBstMidEndSepPunct{\mcitedefaultmidpunct}
{\mcitedefaultendpunct}{\mcitedefaultseppunct}\relax
\EndOfBibitem
\bibitem[Ohnishi and Ten-no(2016)Ohnishi, and Ten-no]{Ohnishi_2016}
Ohnishi,~Y.-y.; Ten-no,~S. Explicitly Correlated Frequency-Independent
  Second-Order Green's Function for Accurate Ionization Energies. \emph{J.
  Comput. Chem.} \textbf{2016}, \emph{37}, 2447--2453\relax
\mciteBstWouldAddEndPuncttrue
\mciteSetBstMidEndSepPunct{\mcitedefaultmidpunct}
{\mcitedefaultendpunct}{\mcitedefaultseppunct}\relax
\EndOfBibitem
\bibitem[Johnson \latin{et~al.}(2018)Johnson, Doran, Ten-no, and
  Hirata]{Johnson_2018}
Johnson,~C.~M.; Doran,~A.~E.; Ten-no,~S.~L.; Hirata,~S. Monte Carlo Explicitly
  Correlated Many-Body Green's Function Theory. \emph{J. Chem. Phys.}
  \textbf{2018}, \emph{149}, 174112\relax
\mciteBstWouldAddEndPuncttrue
\mciteSetBstMidEndSepPunct{\mcitedefaultmidpunct}
{\mcitedefaultendpunct}{\mcitedefaultseppunct}\relax
\EndOfBibitem
\bibitem[Pavo{\v s}evi{\'c} \latin{et~al.}(2017)Pavo{\v s}evi{\'c}, Peng,
  Ortiz, and Valeev]{Pavosevic_2017}
Pavo{\v s}evi{\'c},~F.; Peng,~C.; Ortiz,~J.~V.; Valeev,~E.~F. Communication:
  {{Explicitly}} Correlated Formalism for Second-Order Single-Particle
  {{Green}}'s Function. \emph{J. Chem. Phys.} \textbf{2017}, \emph{147},
  121101\relax
\mciteBstWouldAddEndPuncttrue
\mciteSetBstMidEndSepPunct{\mcitedefaultmidpunct}
{\mcitedefaultendpunct}{\mcitedefaultseppunct}\relax
\EndOfBibitem
\bibitem[Teke \latin{et~al.}(2019)Teke, Pavosevic, Peng, and Valeev]{Teke_2019}
Teke,~N.~K.; Pavosevic,~F.; Peng,~C.; Valeev,~E.~F. Explicitly Correlated
  Renormalized Second-Order Green's Function For Accurate Ionization Potentials
  Of Closed-Shell Molecules. \emph{J. Chem. Phys.} \textbf{2019}, \emph{150},
  214103\relax
\mciteBstWouldAddEndPuncttrue
\mciteSetBstMidEndSepPunct{\mcitedefaultmidpunct}
{\mcitedefaultendpunct}{\mcitedefaultseppunct}\relax
\EndOfBibitem
\bibitem[Levy(1979)]{Levy_1979}
Levy,~M. Universal Variational Functionals Of Electron Densities, First-Order
  Density Matrices, And Natural Spin-Orbitals And Solution Of The
  V-Representability Problem. \emph{Proc. Natl. Acad. Sci. U.S.A.}
  \textbf{1979}, \emph{76}, 6062\relax
\mciteBstWouldAddEndPuncttrue
\mciteSetBstMidEndSepPunct{\mcitedefaultmidpunct}
{\mcitedefaultendpunct}{\mcitedefaultseppunct}\relax
\EndOfBibitem
\bibitem[Levy(1982)]{Levy_1982}
Levy,~M. Electron Densities In Search Of Hamiltonians. \emph{Phys. Rev. A}
  \textbf{1982}, \emph{26}, 1200\relax
\mciteBstWouldAddEndPuncttrue
\mciteSetBstMidEndSepPunct{\mcitedefaultmidpunct}
{\mcitedefaultendpunct}{\mcitedefaultseppunct}\relax
\EndOfBibitem
\bibitem[Lieb(1983)]{Lieb_1983}
Lieb,~E.~H. Density Functionals For Coulomb Systems. \emph{Int. J. Quantum
  Chem.} \textbf{1983}, \emph{{24}}, 243\relax
\mciteBstWouldAddEndPuncttrue
\mciteSetBstMidEndSepPunct{\mcitedefaultmidpunct}
{\mcitedefaultendpunct}{\mcitedefaultseppunct}\relax
\EndOfBibitem
\bibitem[Dahlen and {van Leeuwen}(2005)Dahlen, and {van Leeuwen}]{Dahlen_2005}
Dahlen,~N.~E.; {van Leeuwen},~R. Self-Consistent Solution of the {{Dyson}}
  Equation for Atoms and Molecules within a Conserving Approximation. \emph{J.
  Chem. Phys.} \textbf{2005}, \emph{122}, 164102\relax
\mciteBstWouldAddEndPuncttrue
\mciteSetBstMidEndSepPunct{\mcitedefaultmidpunct}
{\mcitedefaultendpunct}{\mcitedefaultseppunct}\relax
\EndOfBibitem
\bibitem[Dahlen \latin{et~al.}(2005)Dahlen, Van~Leeuwen, and
  Von~Barth]{Dahlen_2005a}
Dahlen,~N.~E.; Van~Leeuwen,~R.; Von~Barth,~U. Variational Energy Functionals of
  the {{Green}} Function Tested on Molecules. \emph{Int. J. Quantum Chem.}
  \textbf{2005}, \emph{101}, 512--519\relax
\mciteBstWouldAddEndPuncttrue
\mciteSetBstMidEndSepPunct{\mcitedefaultmidpunct}
{\mcitedefaultendpunct}{\mcitedefaultseppunct}\relax
\EndOfBibitem
\bibitem[Dahlen \latin{et~al.}(2006)Dahlen, {van Leeuwen}, and {von
  Barth}]{Dahlen_2006}
Dahlen,~N.~E.; {van Leeuwen},~R.; {von Barth},~U. Variational Energy
  Functionals of the {{Green}} Function and of the Density Tested on Molecules.
  \emph{Phys. Rev. A} \textbf{2006}, \emph{73}, 012511\relax
\mciteBstWouldAddEndPuncttrue
\mciteSetBstMidEndSepPunct{\mcitedefaultmidpunct}
{\mcitedefaultendpunct}{\mcitedefaultseppunct}\relax
\EndOfBibitem
\bibitem[Gill(1994)]{Gill_1994}
Gill,~P. M.~W. Molecular Integrals Over Gaussian Basis Functions. \emph{Adv.
  Quantum Chem.} \textbf{1994}, \emph{25}, 141--205\relax
\mciteBstWouldAddEndPuncttrue
\mciteSetBstMidEndSepPunct{\mcitedefaultmidpunct}
{\mcitedefaultendpunct}{\mcitedefaultseppunct}\relax
\EndOfBibitem
\bibitem[Casida(1995)]{Casida_1995}
Casida,~M.~E. Generalization of the Optimized-Effective-Potential Model to
  Include Electron Correlation: {{A}} Variational Derivation of the
  {{Sham}}-{{Schl{\"u}ter}} Equation for the Exact Exchange-Correlation
  Potential. \emph{Phys. Rev. A} \textbf{1995}, \emph{51}, 2005--2013\relax
\mciteBstWouldAddEndPuncttrue
\mciteSetBstMidEndSepPunct{\mcitedefaultmidpunct}
{\mcitedefaultendpunct}{\mcitedefaultseppunct}\relax
\EndOfBibitem
\bibitem[Dreuw and Head-Gordon(2005)Dreuw, and Head-Gordon]{Dreuw_2005}
Dreuw,~A.; Head-Gordon,~M. Single-{{Reference}} Ab {{Initio Methods}} for the
  {{Calculation}} of {{Excited States}} of {{Large Molecules}}. \emph{Chem.
  Rev.} \textbf{2005}, \emph{105}, 4009--4037\relax
\mciteBstWouldAddEndPuncttrue
\mciteSetBstMidEndSepPunct{\mcitedefaultmidpunct}
{\mcitedefaultendpunct}{\mcitedefaultseppunct}\relax
\EndOfBibitem
\bibitem[Veril \latin{et~al.}(2018)Veril, Romaniello, Berger, and
  Loos]{Veril_2018}
Veril,~M.; Romaniello,~P.; Berger,~J.~A.; Loos,~P.~F. Unphysical
  Discontinuities in GW Methods. \emph{J. Chem. Theory Comput.} \textbf{2018},
  \emph{14}, 5220\relax
\mciteBstWouldAddEndPuncttrue
\mciteSetBstMidEndSepPunct{\mcitedefaultmidpunct}
{\mcitedefaultendpunct}{\mcitedefaultseppunct}\relax
\EndOfBibitem
\bibitem[Martin and Schwinger(1959)Martin, and Schwinger]{Martin_1959}
Martin,~P.~C.; Schwinger,~J. Theory of {{Many}}-{{Particle Systems}}. {{I}}.
  \emph{Phys. Rev.} \textbf{1959}, \emph{115}, 1342--1373\relax
\mciteBstWouldAddEndPuncttrue
\mciteSetBstMidEndSepPunct{\mcitedefaultmidpunct}
{\mcitedefaultendpunct}{\mcitedefaultseppunct}\relax
\EndOfBibitem
\bibitem[Baym and Kadanoff(1961)Baym, and Kadanoff]{Baym_1961}
Baym,~G.; Kadanoff,~L.~P. Conservation {{Laws}} and {{Correlation Functions}}.
  \emph{Phys. Rev.} \textbf{1961}, \emph{124}, 287--299\relax
\mciteBstWouldAddEndPuncttrue
\mciteSetBstMidEndSepPunct{\mcitedefaultmidpunct}
{\mcitedefaultendpunct}{\mcitedefaultseppunct}\relax
\EndOfBibitem
\bibitem[Baym(1962)]{Baym_1962}
Baym,~G. Self-{{Consistent Approximations}} in {{Many}}-{{Body Systems}}.
  \emph{Phys. Rev.} \textbf{1962}, \emph{127}, 1391--1401\relax
\mciteBstWouldAddEndPuncttrue
\mciteSetBstMidEndSepPunct{\mcitedefaultmidpunct}
{\mcitedefaultendpunct}{\mcitedefaultseppunct}\relax
\EndOfBibitem
\bibitem[{von Barth} and Holm(1996){von Barth}, and Holm]{vonBarth_1996}
{von Barth},~U.; Holm,~B. Self-Consistent {{GW}} 0 Results for the Electron
  Gas: {{Fixed}} Screened Potential {{W}} 0 within the Random-Phase
  Approximation. \emph{Phys. Rev. B} \textbf{1996}, \emph{54}, 8411\relax
\mciteBstWouldAddEndPuncttrue
\mciteSetBstMidEndSepPunct{\mcitedefaultmidpunct}
{\mcitedefaultendpunct}{\mcitedefaultseppunct}\relax
\EndOfBibitem
\bibitem[Toulouse \latin{et~al.}(2005)Toulouse, Gori-Giorgi, and
  Savin]{Toulouse_2005}
Toulouse,~J.; Gori-Giorgi,~P.; Savin,~A. A Short-Range Correlation Energy
  Density Functional With Multi-Determinantal Reference. \emph{Theor. Chem.
  Acc.} \textbf{2005}, \emph{114}, 305\relax
\mciteBstWouldAddEndPuncttrue
\mciteSetBstMidEndSepPunct{\mcitedefaultmidpunct}
{\mcitedefaultendpunct}{\mcitedefaultseppunct}\relax
\EndOfBibitem
\bibitem[Paziani \latin{et~al.}(2006)Paziani, Moroni, Gori-Giorgi, and
  Bachelet]{Paziani_2006}
Paziani,~S.; Moroni,~S.; Gori-Giorgi,~P.; Bachelet,~G.~B. Local-Spin-Density
  Functional For Multideterminant Density Functional Theory. \emph{Phys. Rev.
  B} \textbf{2006}, \emph{73}, 155111\relax
\mciteBstWouldAddEndPuncttrue
\mciteSetBstMidEndSepPunct{\mcitedefaultmidpunct}
{\mcitedefaultendpunct}{\mcitedefaultseppunct}\relax
\EndOfBibitem
\bibitem[Loos and Gill(2016)Loos, and Gill]{Loos_2016}
Loos,~P.-F.; Gill,~P. M.~W. {{The}} Uniform Electron Gas. \emph{Wiley
  Interdiscip. Rev. Comput. Mol. Sci.} \textbf{2016}, \emph{6}, 410--429\relax
\mciteBstWouldAddEndPuncttrue
\mciteSetBstMidEndSepPunct{\mcitedefaultmidpunct}
{\mcitedefaultendpunct}{\mcitedefaultseppunct}\relax
\EndOfBibitem
\bibitem[Fert\'e \latin{et~al.}(2019)Fert\'e, Giner, and Toulouse]{Ferte_2019}
Fert\'e,~A.; Giner,~E.; Toulouse,~J. Range-Separated Multideterminant
  Density-Functional Theory With A Short-Range Correlation Functional Of The
  On-Top Pair Density. \emph{J. Chem. Phys.} \textbf{2019}, \emph{150},
  084103\relax
\mciteBstWouldAddEndPuncttrue
\mciteSetBstMidEndSepPunct{\mcitedefaultmidpunct}
{\mcitedefaultendpunct}{\mcitedefaultseppunct}\relax
\EndOfBibitem
\bibitem[Perdew \latin{et~al.}(1996)Perdew, Burke, and Ernzerhof]{Perdew_1996}
Perdew,~J.~P.; Burke,~K.; Ernzerhof,~M. Generalized Gradient Approximation Made
  Simple. \emph{Phys. Rev. Lett.} \textbf{1996}, \emph{77}, 3865--3868\relax
\mciteBstWouldAddEndPuncttrue
\mciteSetBstMidEndSepPunct{\mcitedefaultmidpunct}
{\mcitedefaultendpunct}{\mcitedefaultseppunct}\relax
\EndOfBibitem
\bibitem[Toulouse \latin{et~al.}(2004)Toulouse, Colonna, and
  Savin]{Toulouse_2004}
Toulouse,~J.; Colonna,~F.; Savin,~A. Long-Range--Short-Range Separation Of The
  Electron-Electron Interaction In Density-Functional Theory. \emph{Phys. Rev.
  A} \textbf{2004}, \emph{70}, 062505\relax
\mciteBstWouldAddEndPuncttrue
\mciteSetBstMidEndSepPunct{\mcitedefaultmidpunct}
{\mcitedefaultendpunct}{\mcitedefaultseppunct}\relax
\EndOfBibitem
\bibitem[Gori-Giorgi and Savin(2006)Gori-Giorgi, and Savin]{Gori-Giorgi_2006}
Gori-Giorgi,~P.; Savin,~A. Properties Of Short-Range And Long-Range Correlation
  Energy Density Functionals From Electron-Electron Coalescence. \emph{Phys.
  Rev. A} \textbf{2006}, \emph{73}, 032506\relax
\mciteBstWouldAddEndPuncttrue
\mciteSetBstMidEndSepPunct{\mcitedefaultmidpunct}
{\mcitedefaultendpunct}{\mcitedefaultseppunct}\relax
\EndOfBibitem
\bibitem[Garniron \latin{et~al.}(2019)Garniron, Gasperich, Applencourt, Benali,
  Fert{\'e}, Paquier, Pradines, Assaraf, Reinhardt, Toulouse, Barbaresco,
  Renon, David, Malrieu, V{\'e}ril, Caffarel, Loos, Giner, and Scemama]{QP2}
Garniron,~Y.; Gasperich,~K.; Applencourt,~T.; Benali,~A.; Fert{\'e},~A.;
  Paquier,~J.; Pradines,~B.; Assaraf,~R.; Reinhardt,~P.; Toulouse,~J.;
  Barbaresco,~P.; Renon,~N.; David,~G.; Malrieu,~J.~P.; V{\'e}ril,~M.;
  Caffarel,~M.; Loos,~P.~F.; Giner,~E.; Scemama,~A. Quantum Package 2.0: A
  Open-Source Determinant-Driven Suite Of Programs. \emph{J. Chem. Theory
  Comput.} \textbf{2019}, \emph{15}, 3591\relax
\mciteBstWouldAddEndPuncttrue
\mciteSetBstMidEndSepPunct{\mcitedefaultmidpunct}
{\mcitedefaultendpunct}{\mcitedefaultseppunct}\relax
\EndOfBibitem
\bibitem[Duchemin \latin{et~al.}(2017)Duchemin, Li, and Blase]{Duchemin_2017}
Duchemin,~I.; Li,~J.; Blase,~X. Hybrid and Constrained Resolution-of-Identity
  Techniques for Coulomb Integrals. \emph{J. Chem. Theory Comput.}
  \textbf{2017}, \emph{13}, 1199\relax
\mciteBstWouldAddEndPuncttrue
\mciteSetBstMidEndSepPunct{\mcitedefaultmidpunct}
{\mcitedefaultendpunct}{\mcitedefaultseppunct}\relax
\EndOfBibitem
\bibitem[Rojas \latin{et~al.}(1995)Rojas, Godby, and Needs]{Rojas_1995}
Rojas,~H.~N.; Godby,~R.~W.; Needs,~R.~J. Space-Time Method for Ab Initio
  Calculations of Self-Energies and Dielectric Response Functions of Solids.
  \emph{Phys. Rev. Lett.} \textbf{1995}, \emph{74}, 1827\relax
\mciteBstWouldAddEndPuncttrue
\mciteSetBstMidEndSepPunct{\mcitedefaultmidpunct}
{\mcitedefaultendpunct}{\mcitedefaultseppunct}\relax
\EndOfBibitem
\bibitem[Duchemin and Blase(2019)Duchemin, and Blase]{Duchemin_2019}
Duchemin,~I.; Blase,~X. Separable Resolution-of-the-Identity with All-Electron
  Gaussian Bases: Application to Cubic-scaling RPA. \emph{J. Chem. Phys.}
  \textbf{2019}, \emph{150}, 174120\relax
\mciteBstWouldAddEndPuncttrue
\mciteSetBstMidEndSepPunct{\mcitedefaultmidpunct}
{\mcitedefaultendpunct}{\mcitedefaultseppunct}\relax
\EndOfBibitem
\bibitem[Dunning(1989)]{Dunning_1989}
Dunning,~T.~H. Gaussian Basis Sets For Use In Correlated Molecular
  Calculations. I. The Atoms Boron Through Neon And Hydrogen. \emph{J. Chem.
  Phys.} \textbf{1989}, \emph{90}, 1007\relax
\mciteBstWouldAddEndPuncttrue
\mciteSetBstMidEndSepPunct{\mcitedefaultmidpunct}
{\mcitedefaultendpunct}{\mcitedefaultseppunct}\relax
\EndOfBibitem
\bibitem[Lewis and Berkelbach(2019)Lewis, and Berkelbach]{Lewis_2019a}
Lewis,~A.~M.; Berkelbach,~T.~C. Vertex Corrections to the Polarizability Do Not
  Improve the GW Approximation for the Ionization Potential of Molecules.
  \emph{J. Chem. Theory Comput.} \textbf{2019}, \emph{15}, 2925\relax
\mciteBstWouldAddEndPuncttrue
\mciteSetBstMidEndSepPunct{\mcitedefaultmidpunct}
{\mcitedefaultendpunct}{\mcitedefaultseppunct}\relax
\EndOfBibitem
\bibitem[Kim \latin{et~al.}(2013)Kim, Sim, and Burke]{Kim_2013}
Kim,~M.; Sim,~E.; Burke,~K. Understanding and Reducing Errors in Density
  Functional Calculations. \emph{Phys. Rev. Lett.} \textbf{2013}, \emph{111},
  073003\relax
\mciteBstWouldAddEndPuncttrue
\mciteSetBstMidEndSepPunct{\mcitedefaultmidpunct}
{\mcitedefaultendpunct}{\mcitedefaultseppunct}\relax
\EndOfBibitem
\bibitem[Adler \latin{et~al.}(2007)Adler, Knizia, and Werner]{Adler_2007}
Adler,~T.~B.; Knizia,~G.; Werner,~H.-J. A Simple and Efficient {{CCSD(T)-F12}}
  Approximation. \emph{J. Chem. Phys.} \textbf{2007}, \emph{127}, 221106\relax
\mciteBstWouldAddEndPuncttrue
\mciteSetBstMidEndSepPunct{\mcitedefaultmidpunct}
{\mcitedefaultendpunct}{\mcitedefaultseppunct}\relax
\EndOfBibitem
\bibitem[Govoni and Galli(2018)Govoni, and Galli]{Govoni_2018}
Govoni,~M.; Galli,~G. GW100: Comparison of Methods and Accuracy of Results
  Obtained with the WEST Code. \emph{J. Chem. Theory Comput.} \textbf{2018},
  \emph{14}, 1895--1909\relax
\mciteBstWouldAddEndPuncttrue
\mciteSetBstMidEndSepPunct{\mcitedefaultmidpunct}
{\mcitedefaultendpunct}{\mcitedefaultseppunct}\relax
\EndOfBibitem
\bibitem[{Roca-Sanjuan} \latin{et~al.}(2006){Roca-Sanjuan}, Rubio, Merchan, and
  {Serrano-Andres}]{Roca-Sanjuan_2006}
{Roca-Sanjuan},~D.; Rubio,~M.; Merchan,~M.; {Serrano-Andres},~L. Ab Initio
  Determination of The Ionization Potentials of {{DNA}} And {{RNA}}
  Nucleobases. \emph{J. Chem. Phys.} \textbf{2006}, \emph{125}, 084302\relax
\mciteBstWouldAddEndPuncttrue
\mciteSetBstMidEndSepPunct{\mcitedefaultmidpunct}
{\mcitedefaultendpunct}{\mcitedefaultseppunct}\relax
\EndOfBibitem
\bibitem[Krause \latin{et~al.}(2015)Krause, Harding, and Klopper]{Krause_2015}
Krause,~K.; Harding,~M.~E.; Klopper,~W. Coupled-Cluster Reference Values For
  The Gw27 And Gw100 Test Sets For The Assessment Of Gw Methods. \emph{Mol.
  Phys.} \textbf{2015}, \emph{113}, 1952\relax
\mciteBstWouldAddEndPuncttrue
\mciteSetBstMidEndSepPunct{\mcitedefaultmidpunct}
{\mcitedefaultendpunct}{\mcitedefaultseppunct}\relax
\EndOfBibitem
\bibitem[Weigend \latin{et~al.}(2003)Weigend, Furche, and
  Ahlrichs]{Weigend_2003a}
Weigend,~F.; Furche,~F.; Ahlrichs,~R. Gaussian basis sets of quadruple zeta
  valence quality for atoms H-Kr. \emph{J. Chem. Phys.} \textbf{2003},
  \emph{119}\relax
\mciteBstWouldAddEndPuncttrue
\mciteSetBstMidEndSepPunct{\mcitedefaultmidpunct}
{\mcitedefaultendpunct}{\mcitedefaultseppunct}\relax
\EndOfBibitem
\bibitem[Weigend and Ahlrichs(2005)Weigend, and Ahlrichs]{Weigend_2005a}
Weigend,~F.; Ahlrichs,~R. Balanced basis sets of split valence, triple zeta
  valence and quadruple zeta valence quality for H to Rn: Design and assessment
  of accuracy. \emph{Phys. Chem. Chem. Phys.} \textbf{2005}, \emph{7}\relax
\mciteBstWouldAddEndPuncttrue
\mciteSetBstMidEndSepPunct{\mcitedefaultmidpunct}
{\mcitedefaultendpunct}{\mcitedefaultseppunct}\relax
\EndOfBibitem
\bibitem[Strinati(1988)]{Strinati_1988}
Strinati,~G. Application of the {{Green}}'s Functions Method to the Study of
  the Optical Properties of Semiconductors. \emph{Riv. Nuovo Cimento}
  \textbf{1988}, \emph{11}, 1--86\relax
\mciteBstWouldAddEndPuncttrue
\mciteSetBstMidEndSepPunct{\mcitedefaultmidpunct}
{\mcitedefaultendpunct}{\mcitedefaultseppunct}\relax
\EndOfBibitem
\bibitem[Leng \latin{et~al.}(2016)Leng, Jin, Wei, and Ma]{Leng_2016}
Leng,~X.; Jin,~F.; Wei,~M.; Ma,~Y. {{GW}} Method and {{Bethe}}-{{Salpeter}}
  Equation for Calculating Electronic Excitations: {{GW}} Method and
  {{Bethe}}-{{Salpeter}} Equation. \emph{Wiley Interdiscip. Rev. Comput. Mol.
  Sci.} \textbf{2016}, \emph{6}, 532--550\relax
\mciteBstWouldAddEndPuncttrue
\mciteSetBstMidEndSepPunct{\mcitedefaultmidpunct}
{\mcitedefaultendpunct}{\mcitedefaultseppunct}\relax
\EndOfBibitem
\end{mcitethebibliography}

\end{document}